\documentstyle[preprint,aps]{revtex}

\begin{document}
\hoffset-0.5cm

\preprint{}

\title{A Generalized Fluctuation-Dissipation Theorem for Nonlinear
       Response Functions}

\author{Enke Wang\cite{address1} and Ulrich Heinz}

\address{Institut f\"ur Theoretische Physik, Universit\"at Regensburg,\\
   D-93040 Regensburg, Germany}

\date{\today}

\maketitle

\begin{abstract}

A nonlinear generalization of the Fluctuation-Dissipation Theorem (FDT) 
for the $n$-point Green functions and the amputated 1PI vertex functions
at finite temperature is derived in the framework of the Closed Time 
Path formalism. We verify that this generalized FDT coincides with 
known results for $n=2$ and 3. New explicit relations among the 4-point 
nonlinear response and correlation (fluctuation) functions are presented.

\end{abstract} 

\pacs{PACS numbers: }


\section{Introduction}
\label{sec1}

Within the framework of non-equilibrium statistical mechanics a successful 
linear response theory has been established by Kubo \cite{Kubo} and other
authors \cite{Callen,Green}. One of the focuses in linear response theory
is the Fluctuation-Dissipation Theorem (FDT) \cite{Callen} which describes 
the relation between the response and the correlation (fluctuation)
functions. To seek for a nonlinear generalization of the FDT is an 
important and interesting issue 
\cite{Bernard,Peterson,Efremov,Stratonovich,Bochkov,Hao,Chou,Carrington} 
in nonlinear response theory which deals with functions of many time 
arguments. Earlier attempts 
\cite{Bernard,Peterson,Efremov,Stratonovich,Bochkov} to derive a 
nonlinear generalization of the FDT were based on studies of the 
KMS condition \cite{Kubo,Martin}, the conditions of time reversal 
invariance, and the exact algebra of transformations between different 
representations of the thermal Green functions (see Eq.~(\ref{Gtrans}) 
below). As stated in Ref.\cite{Chou}, these attempts failed since none of 
these relations by themselves provide a nonlinear generalization of the 
FDT, and their correct combination has not yet been found.
      
It was shown in Refs.\cite{Hao,Chou} that the Closed Time Path (CTP) 
formalism suggested by Schwinger\cite{Schwinger} and further
elaborated by Keldysh\cite{Keldysh} provides a suitable theoretical 
framework for nonlinear response theory near thermal equilibrium.
By combining the three types of relations mentioned above, the authors 
of Ref.\cite{Chou} arrived at a plausible nonlinear generalization of the 
FDT for the nonlinear response functions, without, however, giving an
explicit proof of its general validity. Their suggested nonlinear 
generalization of the FDT (Eq.~(5.60) in Ref.\cite{Chou}) involves 
a parameter $\varepsilon=\pm 1$ which depends on the signature under 
time reversal of the physical quantity considered. For arbitrary field 
operators it is not clear which value of $\varepsilon$ should be taken, 
which makes it difficult to verify the correctness of the generalized 
FDT given in \cite{Chou}. In this paper we will give a complete derivation
of the generalized FDT for nonlinear response functions. Instead of 
time reversal we exploit the closely related ``tilde conjugation'' 
transformation. We find that the correct generalized FDT equations
contain no sign parameter $\varepsilon$.

Several recent investigations
\cite{Carrington,Frenkel,Braaten,Baier,Evans,Kobes,Aurenche,Eijck,Taylor,Henning,Hou} 
of $n$-point functions $(n\geq 3)$ and their spectral representations
have tried to clarify the analytical continuation from the imaginary 
time formalism (ITF)\cite{Matsubara} to the real time formalism (RTF)
\cite{Schwinger,Keldysh,Henning,Umezawa}, the RTF Feyman rules and 
their transformation between various RTF representations, and the 
calculation of the temperature dependent running coupling constant 
in QCD. The generalized FDT plays a crucial role in the determination 
of spectral representations for the retarded and advanced $n$-point
Green functions in the CTP formalism \cite{Carrington,Aurenche,Eijck,Hou}.
In Ref. \cite{Carrington} the corresponding relations were derived 
for the retarded and advanced 3-point functions, by way of a rather
tedious evaluation of the corresponding Lehmann spectral representation.
As we will show, these same relations can be deduced very easily from 
the generalized FDT which, however, applies equally well to more general 
situations, $n\geq 4$. Due to the known intimate relation between 
transport coefficients and higher order $n$-point Green functions as 
provided by the Kubo formalism \cite{Kubo,Hosoya,Jeon,Wang}, we expect
the generalized FDT to become a useful tool for the calculation of
collective phenomena in, say, a quark-gluon plasma. As a non-trivial 
example for its application we give explicit expressions for the 
relations between the 4-point nonlinear response and correlation 
functions which may be useful for a RTF calculation of the shear 
viscosity in scalar field theories \cite{Hosoya,Jeon,Wang}.

This paper is organized as follows: In Sec.~\ref{sec2} we review some
important relations between various components of the $n$-point RTF Green 
functions in the CTP formalism and the general expressions for the
nonlinear response functions which will be used in the following sections.
In Sec.~\ref{sec3} we derive the generalized FDT for nonlinear response
functions. By solving the generalized FDT equations, explicit relations 
among the correlation (fluctuation) and response functions are obtained
for the cases of $n=2,3,4$ and checked against known results.
In Sec.~\ref{sec4} we repeat the derivation for the amputated 1PI 
vertex functions. A short summary is given in Sec.~\ref{sec5}.
In the appendix we deduce relations between the vertex functions in the 
physical representation and the $R/A$-functions by Aurenche and
Becherrawy~\cite{Aurenche}.

\section{Nonlinear response theory in the real-time formalism}
\label{sec2}

\subsection{$\lowercase{n}$-point Green functions and their generating 
functional}
\label{sec2a}

In the literature there exist three equivalent representations of the RFT
$n$-point Green functions and their generating functional 
\cite{Chou,Chou2}: the ``closed time-path'', ``single-time'' and 
``physical'' representations. 

1. In the closed time-path (CTP) representation, the $n$-point Green 
functions $G_p(1,\dots,n)$ (which may include disconnected parts, 
especially if the fields develop a vacuum condensate) and their
generating functional $Z[J(x)]$ are defined as  
 \begin{eqnarray}
 \label{Gp}
  G_p(1,\dots,n) 
     &\equiv& (-i)^{n-1} \langle T_p[\hat\phi(1)\cdots\hat\phi(n)]\rangle
 \nonumber\\
     &=& i(-1)^n{{\delta^n Z[J(x)]}\over 
               {\delta J(1)\cdots\delta J(n)}}\Bigr|_{J=0}\, ,  
 \\
 \label{Zj}
  Z[J(x)] &\equiv& \Bigl\langle T_p \Bigl[\exp \Bigl(i\int_p 
          d^4x\, J(x)\,{\hat\phi}(x)\Bigr)\Bigr]\Bigr\rangle \, .
 \end{eqnarray}
Throughout this paper the field operator ${\hat\phi}$ will denote
a bosonic (elementary or composite) field; the obvious extension to
fermions is left to the reader. $J(x)$ is a classical external source,
and $\langle\dots\rangle$ the thermal expectation value. The numbers
$1,\dots,n$ stand for the coordinate arguments $x_1,\dots,x_n$ in
4-dimensional Minkowski space, and $p$ denotes the integration
contour (the closed time-path) which consists of a branch $C_1$ 
running from negative infinity to positive infinity and another
branch $C_2$ running back from positive infinity to negative infinity.
$T_p$ represents the time ordering operator along this time-path; it 
is the usual time ordering operator $T$ on the branch $C_1$ and the 
anti-chronological time ordering operator $\widetilde T$ on the branch 
$C_2$. 

2. Instead of the closed time-path which covers the real time axis 
twice one can use a single real time parameter if one distinguishes
the external source $J_1$ on the branch $C_1$ from $J_2$ on the branch 
$C_2$, setting them equal (and to zero) only at the end of the 
calculation. Then Eqs.~(\ref{Gp}) and (\ref{Zj}) can be rewritten as
 \begin{eqnarray}
 \label{Ga1an}
   G_{a_1 \dots a_n}(1,\dots,n) &\equiv& (-i)^{n-1} 
   \langle T_p[\hat\phi_{a_1}(1)\cdots\hat\phi_{a_n}(n)]\rangle
 \nonumber\\
     &=& i(-1)^{a_1+\dots+a_n} 
         \left.{{\delta^n Z[J_1(x),J_2(x)]}\over 
               {\delta J_{a_1}(1)\cdots\delta J_{a_n}(n)}}
         \right|_{J_1=J_2=0}\, ,  
 \\
 \label{Zj1j2}
  Z[J_1(x),J_2(x)] &\equiv& \Bigl\langle T_p \Bigl[\exp \Bigl(
       i\int_{-\infty}^{\infty} d^4x\, 
       [J_1(x){\hat\phi}_1(x)-J_2(x){\hat\phi}_2(x)]
       \Bigr)\Bigr]\Bigr\rangle \, .
 \end{eqnarray}
Here the single time parameter varies from $-\infty$ to $\infty$,
and $a_1,a_2,\dots,a_n=1,2$ indicate on which of the two branches,
$C_1$ or $C_2$, the fields are located. The definition (\ref{Ga1an}) 
gives, for example,
 \begin{equation}
 \label{G2211} 
  G_{\underbrace{{\scriptstyle{2\dots 2}}}_m 
     \underbrace{{\scriptstyle{1\dots 1}}}_{n-m}}(1,\dots, n)
     = (-i)^{n-1} \langle 
   {\widetilde T}[\hat\phi(1)\cdots\hat\phi(m)] 
   T[\hat\phi(m+1)\cdots\hat\phi(n)]\rangle\, .
 \end{equation} 
 
3. The ``physical'' or $r/a$ representation of the formalism is defined by 
setting
 \begin{mathletters}
 \label{JAJR}
 \begin{eqnarray}
  J_a(x) &=& J_1(x)-J_2(x)\, ,\qquad 
  J_r(x)={1\over 2}\Bigl(J_1(x)+J_2(x)\Bigr)\, ,
 \label{JAJRa}\\ 
  \hat\phi_a(x) &=& \hat\phi_1(x)-\hat\phi_2(x)\, ,\qquad 
  \hat\phi_r(x)={1\over 2}\Bigl(\hat\phi_1(x)+\hat\phi_2(x)\Bigr)\, ,
 \label{JAJRb} 
 \end{eqnarray}
 \end{mathletters}
Eqs.~(\ref{Ga1an}) and (\ref{Zj1j2}) are then transformed into 
 \begin{eqnarray}
 \label{Gaf1afn}
  G_{\alpha_1\dots\alpha_n}(1,\dots,n) 
     &\equiv& (-i)^{n-1} 2^{n_r-1} \langle 
      T_p[\hat\phi_{\alpha_1}(1)\cdots\hat\phi_{\alpha_n}(n)]\rangle
 \nonumber\\
     &=& i(-1)^n 2^{n_r-1}{{\delta^n Z[J_a(x),J_r(x)]}\over 
               {\delta J_{\bar \alpha_1}(1)\cdots
                \delta J_{\bar \alpha_n}(n)}}\Bigr|_{J_a=J_r=0}\, ,  
 \\
 \label{ZjAjR}
  Z[J_a(x),J_r(x)] &\equiv& \Bigl\langle T_p \Bigl[\exp \Bigl(
       i\int_{-\infty}^{\infty} d^4x\, [J_a(x)\hat\phi_r(x)+
           J_r(x)\hat\phi_a(x)]\Bigr)\Bigr]\Bigr\rangle \, .
 \end{eqnarray}
Here $\alpha_1,\dots,\alpha_n=a,r$, and $n_r$ is the number 
of indices $r$ among $(\alpha_1,\alpha_2,\dots,\alpha_n)$. For
$\alpha_i=a$ one defines ${\bar \alpha_i}=r$, and vice versa.
One shows generally that \cite{Chou,Chou2}
 \begin{equation}
 \label{Gaaa}
   G_{aa\dots a}(1,\dots,n)=0\, .
 \end{equation}
Furthermore, the functions $G_{raa\dots a}$, $G_{ara\dots a}$, 
$G_{aar\dots a}$, $\dots$, with only one index $r$ can be expressed 
\cite{Chou,Chou2} as sums over expectation values of $n{-}1$ nested 
commutators which are just the fully retarded $n$-point
Green functions defined in Ref.~\cite{Lehmann}. $G_{rr\dots r}$ is 
a sum over expectation values of $n-1$ nested anticommutators;
it is the $n$-point correlation function, with no retarded or 
advanced relations among any of its time arguments. The remaining 
components $G_{\alpha_1\dots\alpha_n}(1,\dots,n)$ are 
sums over expectation values of combinations of $n-1$ nested 
commutators and anticommutators, with both retarded and advanced
relations between some of the $n$ time arguments.

The generating functional for $n$-point connected Green functions
in the closed time-path representation is defined as
 \begin{equation}
 \label{Wj}
   W[J(x)]=-i\ln Z[J(x)] \, .
 \end{equation}
In analogy to Eqs.~(\ref{Zj1j2}) and (\ref{ZjAjR}), we can define its
single-time representation $W[J_1(x),J_2(x)]$ and its physical 
representation $W[J_a(x),J_r(x)]$. Replacing in 
Eqs.~(\ref{Gp},\ref{Ga1an},\ref{Gaf1afn}) the generating functional 
$Z$ by $W$ we obtain the connected Green functions $G^c_p(1,2,\dots,n)$, 
$G^c_{a_1 \dots a_n}(1,\dots,n)$ and $G^c_{\alpha_1\dots\alpha_n} 
(1,\dots,n)$ in the three different representations.

Finally, the amputated 1PI $n$-point vertex functions 
$\Gamma^{(n)}_p(1,\dots,n)$ and their generating functional 
$\Gamma[\varphi(x)]$ in the closed time-path representation 
are defined as 
 \begin{eqnarray}
 \label{Gamp}
  \Gamma^{(n)}_p(1,\dots,n) 
     &\equiv& (-1)^n \left.
               {{\delta^n \Gamma[\varphi(x)]}\over 
                {\delta \varphi(1)\cdots\delta 
                 \varphi(n)}}\right|_{\varphi=\langle\hat\phi\rangle}\, ,  
 \\
 \label{Gamphi}
  \Gamma[\varphi(x)] &\equiv& W[J(x)]-\int_p d^4x\, J(x)\varphi(x)
  \, , 
 \end{eqnarray}
with the classical field
 \begin{equation}
 \label{phic}
   \varphi(x)={{\delta W[J(x)]}\over {\delta J(x)}}\, . 
 \end{equation}
In the single-time representation these definitions are rewritten as
 \begin{eqnarray}
 \label{Gama1an}
  \Gamma^{(n)}_{a_1 \dots a_n}(1,\dots,n) 
     &\equiv& (-1)^n \left. 
               {{\delta^n \Gamma[\varphi_1(x),\varphi_2(x)]}\over 
               {\delta \varphi_{a_1}(1)\cdots
                \delta \varphi_{a_n}(n)}}
                \right|_{\varphi_1=\varphi_2=\langle\hat\phi\rangle}\, ,  
 \\
 \label{Gamphi1phi2}
  \Gamma[\varphi_1(x),\varphi_2(x)] &\equiv& 
    W[J_1(x),J_2(x)]-\int_{-\infty}^{\infty} d^4x\, 
         \Bigl(J_1(x)\varphi_1(x)-J_2(x)\varphi_2(x)\Bigr)\, ,
 \end{eqnarray}
with the classical fields on the two branches of the closed time-path
 \begin{equation}
 \label{phic1c2}
   \varphi_1(x)={{\delta W[J_1(x),J_2(x)]}\over {\delta J_1(x)}}\, ,\qquad
   \varphi_2(x)=-{{\delta W[J_1(x),J_2(x)]}\over {\delta J_2(x)}}\, . 
 \end{equation}
The corresponding definitions in the physical representation are 
 \begin{eqnarray}
 \label{Gamaf1afn}
  \Gamma^{(n)}_{\alpha_1 \dots \alpha_n}(1,\dots,n) 
   &\equiv& (-1)^n 2^{n_a-1} \left.
         {{\delta^n \Gamma[\varphi_a(x),\varphi_r(x)]}\over 
          {\delta \varphi_{\alpha_1}(1)\cdots
           \delta \varphi_{\alpha_n}(n)}}
           \right|_{\varphi_a=0,\varphi_r=\langle\hat\phi\rangle}\, ,  
 \\
 \label{Gamaphiaphir}
  \Gamma[\varphi_a(x),\varphi_r(x)] &\equiv& 
    W[J_a(x),J_r(x)]-\int_{-\infty}^{\infty} d^4x\, 
         \Bigl(J_a(x)\varphi_r(x)+J_r(x)\varphi_a(x)\Bigr)\, ,
 \end{eqnarray}
where $n_a$ is the number of indices $a$ among 
($\alpha_1,\dots,\alpha_n$), and
 \begin{equation}
 \label{phicacr}
   \varphi_a(x)=\varphi_1(x)-\varphi_2(x)\, ,\qquad
   \varphi_r(x)={1\over 2}\Bigl(\varphi_1(x)+\varphi_2(x)\Bigr)\, .
 \end{equation}
Corresponding to Eq.(\ref{Gaaa}) we have
 \begin{equation}
 \label{Gamarrr}
   \Gamma^{(n)}_{rr\dots r}(1,\dots,n)=0\, . 
 \end{equation}

The relation between the $r/a$ vertex functions 
$\Gamma^{(n)}_{\alpha_1 \dots \alpha_n}$ and the $R/A$ vertex functions
defined in Ref.~\cite{Aurenche} can be found in Appendix~\ref{appa}. 

In Refs.~\cite{Chou,Keldysh,Chou2} the following transformation law between 
the single-time and physical representations for the $n$-point Green 
functions (including disconnected parts) was established:
 \begin{equation}
 \label{Gtrans}
 G_{a_1 \dots a_n}(1,\dots,n)=
   2^{1-{n\over 2}} G_{\alpha_1\dots\alpha_n}(1,\dots,n)
   Q_{\alpha_1 a_1} \cdots Q_{\alpha_n a_n}\, .
 \end{equation}
Here 
 \begin{equation}
 \label{Q}
   Q_{a 1}=-Q_{a 2}=Q_{r 1}=Q_{r 2}={1\over \sqrt{2}}
 \end{equation}
are the four elements of the orthogonal Keldysh transformation for 2-point
functions \cite{Keldysh}. In the following, a summation over repeated indices
is understood. In a similar way we derive   
 \begin{eqnarray}
 \label{Gctrans}
 G^c_{a_1 \dots a_n}(1,\dots,n) &=&
   2^{1-{n\over 2}} G^c_{\alpha_1\dots\alpha_n}(1,\dots,n)
   Q_{\alpha_1 a_1} \cdots Q_{\alpha_n a_n}\, ,
 \\
 \label{Gamatrans}
 \Gamma^{(n)}_{a_1 \dots a_n}(1,\dots,n) &=&
   2^{1-{n\over 2}}\Gamma^{(n)}_{\alpha_1\dots\alpha_n}(1,\dots,n)
   Q_{\alpha_1 a_1}\cdots Q_{\alpha_n a_n}\, .
 \end{eqnarray}
The transformation laws for the connected $n$-point Green functions and 
the amputated 1PI vertex functions are thus identical to those for the
normal $n$-point Green functions (including disconnected parts), contrary
to a remark made on page 24 in Ref.~\cite{Chou}.

\subsection{Tilde conjugation and KMS condition}
\label{sec2b}

The tilde conjugation operation is defined by reversing the time 
ordering in coordinate space\cite{Umezawa}. For example, the tilde 
conjugate of Eq.~(\ref{G2211}) is 
 \begin{equation}
 \label{Gt2211}
  {\widetilde G}_{\underbrace{{\scriptstyle{2\dots 2}}}_m 
                  \underbrace{{\scriptstyle{1\dots 1}}}_{n-m}}(1, \dots, n)
   =(-i)^{n-1} \langle T[\hat\phi(1)\cdots\hat\phi(m)] 
   {\widetilde T}[\hat\phi(m+1)\cdots\hat\phi(n)]\rangle\, .
 \end{equation} 
In momentum space one can show that tilde conjugation is equivalent to
complex conjugation:
 \begin{mathletters}
 \label{Gtilde}
 \begin{eqnarray}
  {\widetilde G}_{a_1 \dots a_n}(k_1, \dots, k_n) &=&
   (-1)^{n-1}G^*_{a_1 \dots a_n}(k_1, \dots, k_n)\, ,
  \label{Gtildea}\\
  {\widetilde G}_{\alpha_1 \dots \alpha_n}(k_1, \dots, k_n) &=&
   (-1)^{n-1}G^*_{\alpha_1 \dots \alpha_n}(k_1, \dots, k_n)\, .
 \label{Gtildeb}
 \end{eqnarray} 
 \end{mathletters}
The factor $(-1)^{n-1}$ here results from the corresponding factor
$(-i)^{n-1}$ in the definition (\ref{Ga1an}). Energy-momentum 
conservation requires $k_1+k_2+\dots+k_n=0$. The relations (\ref{Gtilde})
carry over without modification for the connected Green functions
$G^c_{a_1 \dots a_n}(k_1, \dots, k_n)$ and 
$G^c_{\alpha_1 \alpha_2\dots \alpha_n}(k_1, k_2, \dots, k_n)$. 
For the amputated 1PI vertex functions we have   
 \begin{mathletters}
 \label{Gamatilde}
 \begin{eqnarray}
  {\widetilde \Gamma}^{(n)}_{a_1 \dots a_n}(k_1, \dots, k_n) &=&
   -\Gamma^{(n)*}_{a_1 \dots a_n}(k_1,\dots, k_n)\, ,
  \label{Gamatildea}\\
  {\widetilde \Gamma}^{(n)}_{\alpha_1 \dots \alpha_n}
          (k_1, \dots, k_n) &=&
   -\Gamma^{(n)*}_{\alpha_1 \dots \alpha_n}
          (k_1,\dots, k_n)\, .
 \label{Gamatildeb}
 \end{eqnarray} 
 \end{mathletters}
Using the KMS condition for periodicity of the Green functions in 
imaginary time \cite{Kubo,Martin} one derives in coordinate space 
 \begin{mathletters}
 \label{KMS}
 \begin{eqnarray}
  {\widetilde G}_{a_1\dots a_n}(1, \dots, n) &=&
   \exp \left( i\beta\sum_{\{i | a_i=2\}}\partial_{t_i} \right)
   G_{{\bar a}_1 \dots {\bar a}_n}(1, \dots, n)\, ,
  \label{KMSa}\\
  {\widetilde \Gamma}^{(n)}_{a_1 \dots a_n}(1, \cdots, n) &=&
   \exp \left( i\beta\sum_{\{i | a_i=2\}}\partial_{t_i} \right)
   \Gamma^{(n)}_{{\bar a}_1 \dots {\bar a}_n}(1, \dots, n)\, ,
  \label{KMSb}
 \end{eqnarray} 
 \end{mathletters}
where $\beta$ is the inverse temperature, and ${\bar a}_i=2,1$ for 
$a_i=1,2$. The connected Green functions $G^c_{a_1 \dots a_n}(1, \dots, n)$ 
satisfy the same relation Eq.~(\ref{KMSa}) as the disconnected ones. 
In momentum space the exponential prefactors become products of simple 
inverse Boltzmann factors $e^{\beta k^0_i}$.
 
\subsection{General expression for nonlinear response functions}
\label{sec2c}

Let us consider a thermal system with global equilibrium initial 
conditions at $t_0=-\infty$ which is driven out of equilibrium by
a real time-dependent external source $J(t)$ which is switched on 
adiabatically. In the closed time-path representation we define the Green
functions for a nonvanishing external source by
 \begin{equation}
 \label{Dcp}
  D_p(1,\dots,n) 
     = i (-1)^n{{\delta^n Z[J(x)]}\over 
               {\delta J(1)\cdots\delta J(n)}}\, .  
 \end{equation}
As shown in Refs.\cite{Hao,Chou}, the observables in nonlinear response 
theory are the correlation functions $D_r$, $D_{rr}$, $D_{rrr}$, 
$\dots$ in the physical representation. These functions can be 
expressed as a Volterra series in powers of the external source $J$, with 
$G_{ra}$, $D_{raa}$, $D_{rra}$, $\dots$ as their coefficients. Taking
$J_1(x)=J_2(x)=J(x)$, the nonlinear response can be expressed as
 \begin{mathletters}
 \label{Drrr}
 \begin{eqnarray}
  & & D_r(1) = G_r(1)-\int d2\, G_{ra}(1,2) J(2)+
        {1\over {2!}}\int d2\, d3\, G_{raa}(1,2,3) J(2)J(3) +\dots\, ,
  \label{Drrra}\\
  & & D_{rr}(1,2) = G_{rr}(1,2)-\int d3\, G_{rra}(1,2,3) J(3)
  \nonumber\\
      & &\qquad\qquad\qquad\qquad\qquad\quad
        +{1\over {2!}}\int d3\, d4\, G_{rraa}(1,2,3,4) J(3)J(4) +\dots\, ,
  \label{Drrrb}\\
  & & D_{rrr}(1,2,3) = G_{rrr}(1,2,3)-\int d4\, G_{rrra}(1,2,3,4) J(4)
  \nonumber\\ 
      & &\qquad\qquad\qquad\qquad\qquad\qquad\quad
         +{1\over {2!}}\int d4\, d5\,G_{rrraa}(1,2,3,4,5) J(4)J(5) +\dots\, .
  \label{Drrrc}
 \end{eqnarray} 
 \end{mathletters}
Here $d$i is a shorthand for $d^4 x_i$. Following 
Refs.~\cite{Bernard,Peterson,Efremov,Stratonovich,Bochkov}, the correlation 
functions $D_r$, $D_{rr}$, $D_{rrr}$, $\dots$ and $G_r$, $G_{rr}$, 
$G_{rrr}$, \dots are called fluctuation functions in the 
non-equilibrium and equilibrium state, respectively. The fully retarded 
Green functions $G_{ra}$, $G_{raa}$, $G_{raaa}$, etc., correspond to 
linear, second-order, third-order, and higher order response functions 
of the averaged physical field $\varphi$ to the external source $J$.
Similarly, $G_{rra}$, $G_{rraa}$, $G_{rrra}$, $\dots$ are the response 
functions of the fluctuations at different order.

\section{Nonlinear generalization of the fluctuation-dissipation 
         theorem for $\lowercase{n}$-point Green functions} 
\label{sec3}

Eqs.~(\ref{KMSa}) show that not all $2^n$ components of the 
$n$-point Green functions $G_{a_1 \dots a_n}$ in the single-time 
representation are independent of each other. Taking also into account
Eq.~(\ref{Gtrans}) with Eq.~(\ref{Gaaa}), one sees that there are at 
most $2^{n-1}-1$ independent components. We will soon see that the same 
is true for the physical representation of the $n$-point Green function,
$G_{\alpha_1 \dots \alpha_n}$. 

As a simple example consider the 2-point Green function. According
to (\ref{Gaaa}) $G_{aa}=0$. The remaining three components are defined 
in coordinate space by
 \begin{mathletters}
 \label{GrGa}
 \begin{eqnarray}
  G_{ra}(1,2) &=& G^{\rm ret}(1,2) \equiv -i\theta (t_1-t_2)
                \langle [\hat\phi(1), \hat\phi(2)]\rangle\, , 
  \label{GrGaa}\\
  G_{ar}(1,2) &=& G^{\rm adv}(1,2) \equiv i\theta (t_2-t_1)
                \langle [\hat\phi(1), \hat\phi(2)]\rangle\, , 
  \label{GrGab}\\
  G_{rr}(1,2) &=& G^{\rm cor}(1,2) \equiv 
              -i\langle \{\hat\phi(1),\hat\phi(2)\} \rangle \, .
  \label{GrGac}
 \end{eqnarray} 
 \end{mathletters}
In momentum space the retarded and advanced 2-point functions,
$G^{\rm ret}$ and $G^{\rm adv}$, are related by complex conjugation,
while the correlation function $G^{\rm cor}$ is related to them by the
well-known fluctuation-dissipation theorem~\cite{Callen}, in our
notation 
 \begin{equation}
 \label{FDT1}
   G_{rr}(k)=\Bigl( 1+2n(k^0) \Bigr)\Bigl( G_{ra}(k)-G_{ar}(k) \Bigr)\, .
 \end{equation}
Here $n(k^0)= [e^{\beta k^0}-1]^{-1}$ is the Bose-Einstein distribution 
function. The FDT thus plays an important role for the analytic 
structure of the thermal 2-point function.

We will now derive similar relations among the $n$-point $(n\geq 3)$ 
retarded and advanced Green functions in the physical representation.
Since, as discussed in the preceding section, these functions are 
related to the higher order fluctuations and nonlinear response 
functions, the relations among them contain the desired nonlinear 
generalization of the FDT.

Starting from the transformation (\ref{Gtrans}) and using 
$Q_{r {\bar a}_i}=Q_{r a_i}$ and $Q_{a {\bar a}_i}=-Q_{a a_i}$ we get
 \begin{eqnarray}
 \label{G+-}
  &&G_{a_1\dots a_n}(1, \dots, n) \pm
  G_{{\bar a}_1 \dots {\bar a}_n}(1, \dots, n)\\
  \nonumber
  &&=2^{1-{n\over 2}}\sum_{\alpha_1,\dots,\alpha_n=a,r}
   \Bigl( 1\pm (-1)^{n_a(\alpha_1,\dots,\alpha_n)} \Bigr)
   G_{\alpha_1 \dots \alpha_n}(1, \dots, n)
   Q_{\alpha_1 a_1} \cdots Q_{\alpha_n a_n}\, ,
 \end{eqnarray} 
where $n_a(\alpha_1,\dots,\alpha_n)$ counts the number
of $a$ indices among $(\alpha_1,\dots,\alpha_n)$. Eqs.~(\ref{G+-}) 
form $2^{n-1}$ independent equations. We now consider separately
the contributions to the sum on the r.h.s. of (\ref{G+-}) with
fixed even and odd numbers of $a$ indices $n^e_a$ and $n^o_a$, and 
define
 \begin{mathletters}
 \label{naeo}
 \begin{eqnarray}
  G_{n^e_a,(a_1 \dots a_n)}(1, \dots, n) 
    &=& 2^{n\over 2} 
       \sum_{\alpha_1,\dots,\alpha_n=a,r \atop
             n_a(\alpha_1,\dots,\alpha_n)=n^e_a}
      G_{\alpha_1 \dots \alpha_n}(1, \dots, n)
      Q_{\alpha_1 a_1} \cdots Q_{\alpha_n a_n} \, ,
  \label{naeo1}\\
  G_{n^o_a,(a_1 \dots a_n)}(1, \dots, n) 
    &=&2^{n\over 2} 
       \sum_{\alpha_1,\dots,\alpha_n=a,r \atop
             n_a(\alpha_1,\dots,\alpha_n)=n^o_a}
      G_{\alpha_1 \dots \alpha_n}(1, \dots, n)
      Q_{\alpha_1 a_1} \cdots Q_{\alpha_n a_n}\, .
  \label{naeo2}
 \end{eqnarray} 
 \end{mathletters}
Due to Eq.~(\ref{Gaaa}) neither $n^e_a$ nor $n^o_a$ can be larger than 
$n-1$. Eq.~(\ref{G+-}) then gives
 \begin{mathletters}
 \label{sumnaeo}
 \begin{eqnarray}
   \sum_{0\leq n^e_a\leq n-1} G_{n^e_a,(a_1 \dots a_n)}(1, \dots, n) 
   &=&2^{n-2} \Bigl[ G_{a_1 \dots a_n}(1, \dots, n) + 
       G_{{\bar a}_1 \dots {\bar a}_n}(1, \dots, n) \Bigr]\, ,
  \label{sumnaeo1}\\
   \sum_{1\leq n^o_a\leq n-1} G_{n^o_a,(a_1 \dots a_n)}(1, \dots, n)
   &=& 2^{n-2} \Bigl[ G_{a_1 \dots a_n}(1, \dots, n) - 
       G_{{\bar a}_1 \dots {\bar a}_n}(1, \dots, n) \Bigr]\, .
  \label{sumnaeo2}
 \end{eqnarray} 
 \end{mathletters}
With the help of the KMS condition (\ref{KMSa}) and some further 
algebra we find
 \begin{mathletters}
 \label{relation}
 \begin{eqnarray}
   && \sum_{0\leq n^e_a\leq n-1}
      \Bigl( G_{n^e_a,(a_1 \dots a_n)}(1, \dots, n) +
       {\widetilde G}_{n^e_a,(a_1 \dots a_n)}(1, \dots, n)\Bigr)
  \nonumber\\
   &&\qquad =\coth\left({A\over 2}\right) \sum_{1\leq n^o_a\leq n-1}
      \Bigl( G_{n^o_a,(a_1 \dots a_n)}(1, \dots, n) +
       {\widetilde G}_{n^o_a,(a_1 \dots a_n)}(1, \dots, n)\Bigr)\, ,
  \label{relation1}\\
   && \sum_{0\leq n^e_a\leq n-1}
      \Bigl( G_{n^e_a,(a_1 \dots a_n)}(1, \dots, n) -
       {\widetilde G}_{n^e_a,(a_1 \dots a_n)}(1, \dots, n)\Bigr)
  \nonumber\\
   &&\qquad =\tanh\left({A\over 2}\right) \sum_{1\leq n^o_a\leq n-1}
      \Bigl( G_{n^o_a,(a_1 \dots a_n)}(1, \dots, n) -
       {\widetilde G}_{n^o_a,(a_1 \dots a_n)}(1, \dots, n)\Bigr)\, ,
  \label{relation2}
 \end{eqnarray} 
 \end{mathletters}
where 
 \begin{equation}
 \label{A}
  A \equiv i\beta\sum_{\{i | a_i=2\}}\partial_{t_i}\, .
 \end{equation}
Going to momentum space and using (\ref{Gtildea}) we can separate the
real and imaginary parts of these equations. For odd values of $n$ we 
obtain
 \begin{mathletters}
 \label{GFDT1}
 \begin{eqnarray}
  && {\rm Re\,}\left[\sum_{0\leq n^e_a\leq n-1}
      G_{n^e_a,(a_1 \dots a_n)}(k_1, \dots, k_n)\right] 
  \nonumber\\
  &&\qquad =\coth\left( {\beta\over 2}\sum_{\{i | a_i=2\}} k^0_i \right)
      {\rm Re\,}\left[ \sum_{1\leq n^o_a\leq n-2}
      G_{n^o_a,(a_1 \dots a_n)}(k_1, \dots, k_n)\right]\, ,
 \label{GFDT1a}\\
  && {\rm Im\,}\left[\sum_{0\leq n^e_a\leq n-1}
      G_{n^e_a,(a_1 \dots a_n)}(k_1, \dots, k_n)\right]
  \nonumber\\
      &&\qquad =\tanh\left( {\beta\over 2}\sum_{\{i | a_i=2\}} k^0_i \right)
      {\rm Im\,}\left[ \sum_{1\leq n^o_a\leq n-2}
      G_{n^o_a,(a_1 \dots a_n)}(k_1, \dots, k_n)\right]\, ,
 \label{GFDT1b}
 \end{eqnarray} 
 \end{mathletters}
while for even $n$ we find
 \begin{mathletters}
 \label{GFDT2}
 \begin{eqnarray}
  &&{\rm Im\, }\left[\sum_{0\leq n^e_a\leq n-2}
      G_{n^e_a,(a_1 \dots a_n)}(k_1, \dots, k_n)\right] 
  \nonumber\\
  &&\qquad =\coth\left( {\beta\over 2}\sum_{\{i | a_i=2\}} k^0_i \right)
      {\rm Im\,}\left[ \sum_{1\leq n^o_a\leq n-1}
      G_{n^o_a,(a_1 \dots a_n)}(k_1, \dots, k_n)\right]\, ,
 \label{GFDT2a}\\
  && {\rm Re\,}\left[\sum_{0\leq n^e_a\leq n-2}
      G_{n^e_a,(a_1 \dots a_n)}(k_1, \dots, k_n)\right]
  \nonumber\\
  &&\qquad =\tanh\left( {\beta\over 2}\sum_{\{i | a_i=2\}} k^0_i \right)
      {\rm Re\,}\left[ \sum_{1\leq n^o_a\leq n-1}
      G_{n^o_a,(a_1 \dots a_n)}(k_1, \dots, k_n)\right] \, .
 \label{GFDT2b}
 \end{eqnarray} 
 \end{mathletters}
Eqs.~(\ref{GFDT1},\ref{GFDT2}) are the nonlinear generalization of the 
FDT for $n$-point Green functions. Each of these two sets of equations
contains $2^{n-1}$ pairs of (real) relations among the components
$G_{\alpha_1 \cdots \alpha_n}$ in the physical representation. 
Taking also into account Eq.~(\ref{Gaaa}) we see that at most 
$2^{n-1}-1$ (complex) components are independent, as promised above.

We should point out that the nonlinear generalization of the FDT 
for the connected $n$-point Green functions is the same as 
Eqs.~(\ref{GFDT1},\ref{GFDT2}). This is because $G$ and $G^c$ have
the same transformation properties (\ref{Gtrans}) and (\ref{Gtilde}) 
and satisfy the same KMS condition (\ref{KMSa}). The following examples 
will therefore be written down only for the standard $n$-point functions
which may contain disconnected contributions like, e.g., vacuum
condensates.

\subsection{2-point function}
\label{sec3a}

Combining Eqs.~(\ref{Gaaa}) and (\ref{Gtrans}) for $n=2$ leads to
 \begin{mathletters}
 \label{G2}
 \begin{eqnarray}
   G_{22} &=& {1\over 2}( G_{rr}-G_{ra}-G_{ar} )\, ,\qquad
   G_{21}  =  {1\over 2}( G_{rr}+G_{ra}-G_{ar} )\, ,
  \label{G2a}\\
   G_{12} &=& {1\over 2}( G_{rr}-G_{ra}+G_{ar} )\, ,\qquad
   G_{11}  =  {1\over 2}( G_{rr}+G_{ra}+G_{ar} )\, .
  \label{G2b}
 \end{eqnarray} 
 \end{mathletters}
We can then write down the definitions (\ref{naeo}) for the two relevant
cases $n_a^e=0$, $n_a^o=1$:
 \begin{mathletters}
 \label{G2naeo}
 \begin{eqnarray}
   G_{0,(22)} &=& G_{rr}\, ,\qquad\qquad 
   G_{1,(22)} = -G_{ra}-G_{ar}\, ;
  \label{G2naeo1}\\
   G_{0,(21)} &=& G_{rr}\, ,\qquad\qquad 
   G_{1,(21)} = G_{ra}-G_{ar}\, .
  \label{G2naeo2}
 \end{eqnarray} 
 \end{mathletters}
Substitution into (\ref{GFDT2}) yields in momentum space
 \begin{mathletters}
 \label{G2eq}
 \begin{eqnarray}
   &&{\rm Im}\Bigl[ G_{ra}(k_1, k_2)+G_{ar}(k_1, k_2)\Bigr] = 0\, , 
  \label{G2eq1}\\
   &&{\rm Re\,}G_{rr}(k_1, k_2) = 0\, ; 
  \label{G2eq2}\\
   &&{\rm Im\,}G_{rr}(k_1, k_2) = \coth{{\beta k^0_1}\over 2}\,
      {\rm Im}\Bigl[ G_{ra}(k_1, k_2)-G_{ar}(k_1, k_2)\Bigr]  \, , 
  \label{G2eq3}\\
   &&{\rm Re\,}G_{rr}(k_1, k_2) = \tanh{{\beta k^0_1}\over 2}\,
      {\rm Re}\Bigl[ G_{ra}(k_1, k_2)-G_{ar}(k_1, k_2)\Bigr]  \, . 
  \label{G2eq4}
 \end{eqnarray} 
 \end{mathletters}
The zeros in the first two equations follow from energy conservation, 
$k^0_1+k^0_2=0$, which leads to $\coth[\beta (k^0_1+k^0_2)/2]\to\infty$ 
and $\tanh[\beta (k^0_1+k^0_2)/2]\to 0$. Eqs.~(\ref{G2eq}) are easily 
solved by
 \begin{equation}
 \label{G2sol}
   G_{rr}(k) = \coth{{\beta k^0}\over 2}
      \Bigl[ G_{ra}(k)-G_{ar}(k)\Bigr]\, ,\qquad 
   G_{ra}(k) = G^*_{ar}(k)\, . 
 \end{equation} 
The first of these equations is the well-known fluctuation-dissipation
theorem (\ref{FDT1}). 

Substituting Eq.(\ref{G2sol}) into Eq.(\ref{G2}) and using the
two-component column vectors introduced in
Refs.~\cite{Henning,Umezawa} one can express Eq.~(\ref{G2}) in terms
of an outer product of these column vectors~\cite{Henning,Carrington} 
 \begin{equation}
 \label{G2vector}
   G_{a_1 a_2}(k) = G_{ra}(k){1\choose 1}{{1+n(k^0)}\choose n(k^0)}-
          G_{ar}(k){n(k^0)\choose {1+n(k^0)}}{1\choose 1}\, .
 \end{equation} 

\subsection{3-point function}
\label{sec3b}

For $n=3$ Eqs.~(\ref{Gaaa}) and (\ref{Gtrans}) give
 \begin{mathletters}
 \label{G3}
 \begin{eqnarray}
   G_{222} &=& {1\over 4}( G_{rrr}-G_{rra}-G_{rar}-G_{arr}
                          +G_{raa}+G_{ara}+G_{aar})\, ,
  \label{G3a}\\
   G_{211} &=& {1\over 4}( G_{rrr}+G_{rra}+G_{rar}-G_{arr}
                          +G_{raa}-G_{ara}-G_{aar})\, ,
  \label{G3b}\\
   G_{121} &=& {1\over 4}( G_{rrr}+G_{rra}-G_{rar}+G_{arr}
                          -G_{raa}+G_{ara}-G_{aar})\, ,
  \label{G3c}\\
   G_{112} &=& {1\over 4}( G_{rrr}-G_{rra}+G_{rar}+G_{arr}
                          -G_{raa}-G_{ara}+G_{aar})\, ,
  \label{G3d}\\
   G_{221} &=& {1\over 4}( G_{rrr}+G_{rra}-G_{rar}-G_{arr}
                          -G_{raa}-G_{ara}+G_{aar})\, ,
  \label{G3e}\\
   G_{212} &=& {1\over 4}( G_{rrr}-G_{rra}+G_{rar}-G_{arr}
                          -G_{raa}+G_{ara}-G_{aar})\, ,
  \label{G3f}\\
   G_{122} &=& {1\over 4}( G_{rrr}-G_{rra}-G_{rar}+G_{arr}
                          +G_{raa}-G_{ara}-G_{aar})\, ,
  \label{G3g}\\
   G_{111} &=& {1\over 4}( G_{rrr}+G_{rra}+G_{rar}+G_{arr}
                          +G_{raa}+G_{ara}+G_{aar})\, .
  \label{G3h}
 \end{eqnarray} 
 \end{mathletters}
The definitions (\ref{naeo}) yield 
 \begin{mathletters}
 \label{G3naeo}
 \begin{eqnarray}
   && G_{0,(222)} = G_{0,(211)} = G_{0,(121)} = G_{0,(112)} = G_{rrr}\, ;
  \label{G3naeo1}\\
   && G_{1,(222)} = -G_{rra}-G_{rar}-G_{arr}\, ,\qquad
      G_{2,(222)} = G_{raa}+G_{ara}+G_{aar}\, ;
  \label{G3naeo2}\\
   && G_{1,(211)} = G_{rra}+G_{rar}-G_{arr}\, ,\qquad
      G_{2,(211)} = G_{raa}-G_{ara}-G_{aar}\, ;
  \label{G3naeo3}\\
   && G_{1,(121)} = G_{rra}-G_{rar}+G_{arr}\, ,\qquad
      G_{2,(121)} = -G_{raa}+G_{ara}-G_{aar}\, ;
  \label{G3naeo4}\\
   && G_{1,(112)} = -G_{rra}+G_{rar}+G_{arr}\, ,\qquad
      G_{2,(112)} = -G_{raa}-G_{ara}+G_{aar}\, .
  \label{G3naeo5}
 \end{eqnarray} 
 \end{mathletters}
Substituting these quantities into Eqs.~(\ref{GFDT1}) we get in
momentum space
 \begin{mathletters}
 \label{G3eq}
 \begin{eqnarray}
   &&{\rm Re}\Bigl[ G_{rra}+G_{rar}+G_{arr}\Bigr] = 0\, , 
  \label{G3eq1}\\
   &&{\rm Im}\Bigl[ G_{rrr}+G_{raa}+G_{ara}+G_{aar}\Bigr] = 0\, ; 
  \label{G3eq2}\\
   &&{\rm Re}\Bigl[ G_{rrr}+G_{raa}-G_{ara}-G_{aar}\Bigr] = 
     \coth{{\beta k^0_1}\over 2}\,
     {\rm Re}\Bigl[ G_{rra}+G_{rar}-G_{arr}\Bigr]\, ,
  \label{G3eq3}\\
   &&{\rm Im}\Bigl[ G_{rrr}+G_{raa}-G_{ara}-G_{aar}\Bigr] = 
     \tanh{{\beta k^0_1}\over 2}\,
     {\rm Im}\Bigl[ G_{rra}+G_{rar}-G_{arr}\Bigr]\, ;
  \label{G3eq4}\\
   &&{\rm Re}\Bigl[ G_{rrr}-G_{raa}+G_{ara}-G_{aar}\Bigr] = 
     \coth{{\beta k^0_2}\over 2}\,
     {\rm Re}\Bigl[ G_{rra}-G_{rar}+G_{arr}\Bigr]\, ,
  \label{G3eq5}\\
   &&{\rm Im}\Bigl[ G_{rrr}-G_{raa}+G_{ara}-G_{aar}\Bigr] = 
     \tanh{{\beta k^0_2}\over 2}\,
     {\rm Im}\Bigl[ G_{rra}-G_{rar}+G_{arr}\Bigr]\, ;
  \label{G3eq6}\\
   &&{\rm Re}\Bigl[ G_{rrr}-G_{raa}-G_{ara}+G_{aar}\Bigr] = 
     \coth{{\beta k^0_3}\over 2}\,
     {\rm Re}\Bigl[ -G_{rra}+G_{rar}+G_{arr}\Bigr]\, ,
  \label{G3eq7}\\
   &&{\rm Im}\Bigl[ G_{rrr}-G_{raa}-G_{ara}+G_{aar}\Bigr] = 
     \tanh{{\beta k^0_3}\over 2}\,
     {\rm Im}\Bigl[ -G_{rra}+G_{rar}+G_{arr}\Bigr]\, .
  \label{G3eq8}
 \end{eqnarray} 
 \end{mathletters}
Again the zeros in the first two equations result from energy 
conservation. To solve Eqs.~(\ref{G3eq}), we choose $G_{raa}$, 
$G_{ara}$ and $G_{aar}$ as the $2^{n-1}-1=3$ independent physical 
components; in the notation of Ref.~\cite{Carrington} these are the 
three retarded 3-point functions $\Gamma_{Ri}$, $\Gamma_{R}$, and 
$\Gamma_{Ro}$, respectively. We further introduce the shorthands 
\cite{Carrington} $N_i=N(k^0_i)=\coth(\beta k^0_i/2)=1+2n(k^0_i)=1+2n_i$ 
which satisfy the identity
 \begin{equation}
 \label{N}
  N_1 N_2 +N_2 N_3 +N_3 N_1=-1 \, .
 \end{equation}
Solving Eqs.~(\ref{G3eq}) in terms of the three selected components 
we find
 \begin{mathletters}
 \label{G3sol}
 \begin{eqnarray}
   G_{rar} &=& N_1 ( G^*_{ara}-G_{aar} )+
               N_3 ( G^*_{ara}-G_{raa} )\, , 
  \label{G3sol1}\\
   G_{arr} &=& N_2 ( G^*_{raa}-G_{aar} )+
               N_3 ( G^*_{raa}-G_{ara} )\, , 
  \label{G3sol2}\\
   G_{rra} &=& N_1 ( G^*_{aar}-G_{ara} )+
               N_2 ( G^*_{aar}-G_{raa} )\, , 
  \label{G3sol3}\\
   G_{rrr} &=& G^*_{raa}+ G^*_{ara}+ G^*_{aar} 
              +N_2 N_3 ( G_{raa}+G^*_{raa} )
  \nonumber\\
           & &+N_1 N_3 ( G_{ara}+G^*_{ara} ) 
              +N_1 N_2 ( G_{aar}+G^*_{aar} )\, .
  \label{G3sol4}
 \end{eqnarray} 
 \end{mathletters}
Noting the identities $G_{rar}=\Gamma_F$, $G_{arr}=\Gamma_{Fi}$,
$G_{rra}=\Gamma_{Fo}$, and $G_{rrr}=\Gamma_E$ between the 3-point
functions in $r/a$ notation and the 3-point functions used in 
\cite{Carrington}, Eq.~(\ref{G3sol}) are seen to agree with
Eqs.~(33) in \cite{Carrington}. Obviously, the algebra leading from
(\ref{G3eq}) to (\ref{G3sol}) is much simpler than the method of
derivation presented in \cite{Carrington}.

Eq.~(\ref{G3sol4}) expresses the fluctuation function $G_{rrr}$ in 
terms of the second order non-linear response functions 
($G_{raa}$, $G_{ara}$, $G_{aar}$), i.e. in terms of the fully 
retarded vertex functions. Furthermore, Eqs.~(\ref{G3sol}a-c) express 
the linear response functions for the fluctuation 
($G_{rra}$, $G_{rar}$, $G_{arr}$) in terms of the same set of 
retarded Green functions. These four relations form the nonlinear 
generalization of the FDT for the case $n=3$.

When substituting Eq.~(\ref{G3sol}) into (\ref{G3}), Eq.(\ref{G3}) can
be written in terms of outer products of column vectors
as~\cite{Carrington}     
 \begin{eqnarray}
 \label{G3vector}
   &&\Bigl(G_{a_1 a_2 a_3}\Bigr) =
    \nonumber\\ 
    &&\qquad G_{raa}{1\choose 1}{n_2\choose {1+n_2}}{n_3\choose {1+n_3}}
         -G^*_{raa}\Bigl[(1+n_2)(1+n_3)-n_2 n_3\Bigr]
         {n_1\choose {1+n_1}}{1\choose 1}{1\choose 1}
     \nonumber\\
    &&\quad + G_{ara}{n_1\choose {1+n_1}}{1\choose 1}{n_3\choose {1+n_3}}
         -G^*_{ara}\Bigl[(1+n_1)(1+n_3)-n_1 n_3\Bigr]
         {1\choose 1}{n_2\choose {1+n_2}}{1\choose 1}
     \nonumber\\
    &&\quad + G_{aar}{n_1\choose {1+n_1}}{n_2\choose {1+n_2}}{1\choose 1}
         -G^*_{aar}\Bigl[(1+n_1)(1+n_2)-n_1 n_2\Bigr]
         {1\choose 1}{1\choose 1}{n_3\choose {1+n_3}}
 \end{eqnarray} 

\subsection{4-point function}
\label{sec3c}

Following the same steps as for $n=2,3$, we get from Eqs.~(\ref{Gaaa}),
(\ref{Gtrans}), and (\ref{naeo}) 
 \begin{mathletters}
 \label{G4naeo}
 \begin{eqnarray}
   G_{0,(2222)} &=& G_{0,(2221)} = G_{0,(2212)} = G_{0,(2122)} = G_{0,(1222)}
   \nonumber\\
      &=& G_{0,(2211)} = G_{0,(2121)} = G_{0,(1221)}=G_{rrrr}\, ;
  \label{G4naeo1}\\
   G_{1,(2222)} &=& -G_{rrra}-G_{rrar}-G_{rarr}-G_{arrr}\, , 
   \nonumber\\
   G_{2,(2222)} &=& G_{rraa}+G_{rara}+G_{raar}+G_{arra}+G_{arar}+G_{aarr}\, ,
  \nonumber\\
   G_{3,(2222)} &=& -G_{raaa}-G_{araa}-G_{aara}-G_{aaar}\, ;
  \label{G4naeo2}\\
   G_{1,(2221)} &=& G_{rrra}-G_{rrar}-G_{rarr}-G_{arrr}\, , 
   \nonumber\\
   G_{2,(2221)} &=& -G_{rraa}-G_{rara}+G_{raar}-G_{arra}+G_{arar}+G_{aarr}\, ,
  \nonumber\\
   G_{3,(2221)} &=& G_{raaa}+G_{araa}+G_{aara}-G_{aaar}\, ;
  \label{G4naeo3}\\
   G_{1,(2212)} &=& -G_{rrra}+G_{rrar}-G_{rarr}-G_{arrr}\, , 
   \nonumber\\
   G_{2,(2212)} &=& -G_{rraa}+G_{rara}-G_{raar}+G_{arra}-G_{arar}+G_{aarr}\, ,
  \nonumber\\
   G_{3,(2212)} &=&  G_{raaa}+G_{araa}-G_{aara}+G_{aaar}\, ;
  \label{G4naeo4}\\
   G_{1,(2122)} &=& -G_{rrra}-G_{rrar}+G_{rarr}-G_{arrr}\, , 
   \nonumber\\
   G_{2,(2122)} &=& G_{rraa}-G_{rara}-G_{raar}+G_{arra}+G_{arar}-G_{aarr}\, ,
  \nonumber\\
   G_{3,(2122)} &=&  G_{raaa}-G_{araa}+G_{aara}+G_{aaar}\, ;
  \label{G4naeo5}\\
   G_{1,(1222)} &=& -G_{rrra}-G_{rrar}-G_{rarr}+G_{arrr}\, , 
   \nonumber\\
   G_{2,(1222)} &=& G_{rraa}+G_{rara}+G_{raar}-G_{arra}-G_{arar}-G_{aarr}\, ,
  \nonumber\\
   G_{3,(1222)} &=& -G_{raaa}+G_{araa}+G_{aara}+G_{aaar}\, ;
  \label{G4naeo6}\\
   G_{1,(2211)} &=&  G_{rrra}+G_{rrar}-G_{rarr}-G_{arrr}\, , 
   \nonumber\\
   G_{2,(2211)} &=& G_{rraa}-G_{rara}-G_{raar}-G_{arra}-G_{arar}+G_{aarr}\, ,
  \nonumber\\
   G_{3,(2211)} &=& -G_{raaa}-G_{araa}+G_{aara}+G_{aaar}\, ;
  \label{G4naeo7}\\
   G_{1,(2121)} &=&  G_{rrra}-G_{rrar}+G_{rarr}-G_{arrr}\, , 
   \nonumber\\
   G_{2,(2121)} &=& -G_{rraa}+G_{rara}-G_{raar}-G_{arra}+G_{arar}-G_{aarr}\, ,
  \nonumber\\
   G_{3,(2121)} &=& -G_{raaa}+G_{araa}-G_{aara}+G_{aaar}\, ;
  \label{G4naeo8}\\
   G_{1,(1221)} &=& G_{rrra}-G_{rrar}-G_{rarr}+G_{arrr}\, , 
   \nonumber\\
   G_{2,(1221)} &=& -G_{rraa}-G_{rara}+G_{raar}+G_{arra}-G_{arar}-G_{aarr}\, ,
  \nonumber\\
   G_{3,(1221)} &=& G_{raaa}-G_{araa}-G_{aara}+G_{aaar}\, .
  \label{G4naeo9}
 \end{eqnarray} 
 \end{mathletters}
Substituting these quantities into Eq.~(\ref{GFDT2}), we get in 
momentum space
 \begin{mathletters}
 \label{G4eq}
 \begin{eqnarray}
   &&{\rm Im}[ G_{1,(2222)}+G_{3,(2222)}]=0\, ,
  \label{G4eq1}\\
   &&{\rm Re}[ G_{0,(2222)}+G_{2,(2222)}]=0\, ;
  \label{G4eq2}\\
   &&{\rm Im}[ G_{0,(2221)}+G_{2,(2221)}]=-\coth{{\beta k^0_4}\over 2}\,
     {\rm Im}[ G_{1,(2221)}+G_{3,(2221)}]\, ,
  \label{G4eq3}\\
   &&{\rm Re}[ G_{0,(2221)}+G_{2,(2221)}]=-\tanh{{\beta k^0_4}\over 2}\,
     {\rm Re}[ G_{1,(2221)}+G_{3,(2221)}]\, ;
  \label{G4eq4}\\
   &&{\rm Im}[ G_{0,(2212)}+G_{2,(2212)}]=-\coth{{\beta k^0_3}\over 2}\,
     {\rm Im}[ G_{1,(2212)}+G_{3,(2212)}]\, ,
  \label{G4eq5}\\
   &&{\rm Re}[ G_{0,(2212)}+G_{2,(2212)}]=-\tanh{{\beta k^0_3}\over 2}\,
     {\rm Re}[ G_{1,(2212)}+G_{3,(2212)}]\, ;
  \label{G4eq6}\\
   &&{\rm Im}[ G_{0,(2122)}+G_{2,(2122)}]=-\coth{{\beta k^0_2}\over 2}\,
     {\rm Im}[ G_{1,(2122)}+G_{3,(2122)}]\, ,
  \label{G4eq7}\\
   &&{\rm Re}[ G_{0,(2122)}+G_{2,(2122)}]=-\tanh{{\beta k^0_2}\over 2}\,
     {\rm Re}[ G_{1,(2122)}+G_{3,(2122)}]\, ;
  \label{G4eq8}\\
   &&{\rm Im}[ G_{0,(1222)}+G_{2,(1222)}]=-\coth{{\beta k^0_1}\over 2}\,
     {\rm Im}[ G_{1,(1222)}+G_{3,(1222)}]\, ,
  \label{G4eq9}\\
   &&{\rm Re}[ G_{0,(1222)}+G_{2,(1222)}]=-\tanh{{\beta k^0_1}\over 2}\,
     {\rm Re}[ G_{1,(1222)}+G_{3,(1222)}]\, ;
  \label{G4eq10}\\
   &&{\rm Im}[ G_{0,(2211)}+G_{2,(2211)}]=\coth{{\beta(k^0_1+k^0_2)}\over 2}\,
     {\rm Im}[ G_{1,(2211)}+G_{3,(2211)}]\, ,
  \label{G4eq11}\\
   &&{\rm Re}[ G_{0,(2211)}+G_{2,(2211)}]=\tanh{{\beta(k^0_1+k^0_2)}\over 2}\,
     {\rm Re}[ G_{1,(2211)}+G_{3,(2211)}]\, ;
  \label{G4eq12}\\
   &&{\rm Im}[ G_{0,(2121)}+G_{2,(2121)}]=\coth{{\beta(k^0_1+k^0_3)}\over 2}\,
     {\rm Im}[ G_{1,(2121)}+G_{3,(2121)}]\, ,
  \label{G4eq13}\\
   &&{\rm Re}[ G_{0,(2121)}+G_{2,(2121)}]=\tanh{{\beta(k^0_1+k^0_3)}\over 2}\,
     {\rm Re}[ G_{1,(2121)}+G_{3,(2121)}]\, ;
  \label{G4eq14}\\
   &&{\rm Im}[ G_{0,(1221)}+G_{2,(1221)}]=\coth{{\beta(k^0_2+k^0_3)}\over 2}\,
     {\rm Im}[ G_{1,(1221)}+G_{3,(1221)}]\, ,
  \label{G4eq15}\\
   &&{\rm Re}[ G_{0,(1221)}+G_{2,(1221)}]=\tanh{{\beta(k^0_2+k^0_3)}\over 2}\,
     {\rm Re}[ G_{1,(1221)}+G_{3,(1221)}]\, .
  \label{G4eq16}
 \end{eqnarray} 
 \end{mathletters}
Once more the zeros in the first two equations follow from energy 
conservation. For $n=4$ there are at most $2^{n-1}-1=7$ independent 
components of $G_{\alpha_1\alpha_2\alpha_3\alpha_4}$; we choose them 
as $G_{raaa}$, $G_{araa}$, $G_{aara}$, $G_{aaar}$, $G_{arra}$, 
$G_{arar}$ and $G_{aarr}$. Introducing the notation
 \begin{equation}
 \label{Nijkl}
  N^{(ij)}_{(kl)}={{N_i+N_j}\over {N_k+N_l}}={{1+n_i+n_j}\over {1+n_k+n_l}}\, ,
 \end{equation}
we can express the solution of the above equations as 
 \begin{mathletters}
 \label{G4sol}
 \begin{eqnarray}
   G_{rrrr} =&& -N_2 N_3 N_4 G_{raaa}
    +\Bigl( N_2 N_3 N_4 +N_2 +N_3 +N_4\Bigr) G^*_{raaa}
    +N_1 N_3 N_4 G_{araa} 
   \nonumber\\
    &&+N_2 \Bigl( N^{(14)}_{(23)}
      +N_4^2 N^{(13)}_{(24)}\Bigr) G^*_{araa}
      +N_1 N_2 N_4 G_{aara} +N_3 \Bigl( N^{(12)}_{(34)}
      +N_2^2 N^{(14)}_{(23)}\Bigr) G^*_{aara}
   \nonumber\\
    &&+N_1 N_2 N_3 G_{aaar} +N_4 \Bigl( N^{(13)}_{(24)}
      +N_3^2 N^{(12)}_{(34)}\Bigr) G^*_{aaar}
      +N_1 N_4 G_{arra}+N_2 N_3 N^{(14)}_{(23)}G^*_{arra} 
   \nonumber\\
    &&+N_1 N_3 G_{arar}
      +N_2 N_4 N^{(13)}_{(24)}G^*_{arar}+N_1 N_2 G_{aarr}
      +N_3 N_4 N^{(12)}_{(34)}G^*_{aarr}\, ,
  \label{G4sol1}\\
   G_{rrra} =&&N_2 N_3G_{raaa}-N_2 N_4 N^{(13)}_{(24)}G^*_{araa} 
      -N_3 N_4 N^{(12)}_{(34)}G^*_{aara}
      -\Bigl( N^{(13)}_{(24)}+N_3^2 N^{(12)}_{(34)}\Bigr) G^*_{aaar}
   \nonumber\\
    &&-N_1 G_{arra}-N_2 N^{(13)}_{(24)}G^*_{arar}
      -N_3 N^{(12)}_{(34)}G^*_{aarr}\, ,
  \label{G4sol2}\\
   G_{rrar} =&&N_2 N_4G_{raaa}-N_2 N_3 N^{(14)}_{(23)}G^*_{araa} 
      -\Bigl( N^{(12)}_{(34)}+N_2^2 N^{(14)}_{(23)}\Bigr) G^*_{aara}
      -N_3 N_4 N^{(12)}_{(34)}G^*_{aaar}
   \nonumber\\
    &&-N_2 N^{(14)}_{(23)}G^*_{arra}-N_1 G_{arar}
      -N_4 N^{(12)}_{(34)}G^*_{aarr}\, ,
  \label{G4sol3}\\
   G_{rarr} =&&N_3 N_4G_{raaa} 
      -\Bigl( N^{(14)}_{(23)}+N_4^2 N^{(13)}_{(24)}\Bigr) G^*_{araa}
      -N_2 N_3 N^{(14)}_{(23)}G^*_{aara}-N_2 N_4 N^{(13)}_{(24)}G^*_{aaar}
   \nonumber\\
    &&-N_3 N^{(14)}_{(23)}G^*_{arra}
      -N_4 N^{(13)}_{(24)}G^*_{arar}-N_1 G_{aarr}\, ,
  \label{G4sol4}\\
   G_{arrr} =&&\Bigl( 1+ N_2 N_3+ N_2 N_4+ N_3 N_4\Bigr)G^*_{raaa} 
      -N_3 N_4 G_{araa}-N_2 N_4 G_{aara}-N_2 N_3 G_{aaar}
   \nonumber\\
    &&-N_4 G_{arra}-N_3 G_{arar}-N_2 G_{aarr}
  \label{G4sol5}\\
   G_{raar} =&& -N_4 G_{raaa}+N_3 N^{(14)}_{(23)}G^*_{araa}
      +N_2 N^{(14)}_{(23)}G^*_{aara}-N_1 G_{aaar}
      +N^{(14)}_{(23)}G^*_{arra}\, , 
  \label{G4sol6}\\
   G_{rara} =&& -N_3 G_{raaa}+N_4 N^{(13)}_{(24)}G^*_{araa}
      -N_1 G_{aara}+N_2 N^{(13)}_{(24)}G^*_{aaar}
      +N^{(13)}_{(24)}G^*_{arar}\, , 
  \label{G4sol7}\\
   G_{rraa} =&& -N_2 G_{raaa}-N_1 G_{araa}+N_4 N^{(12)}_{(34)}G^*_{aara}
      +N_3 N^{(12)}_{(34)}G^*_{aaar}+N^{(12)}_{(34)}G^*_{aarr}\, . 
  \label{G4sol8}
 \end{eqnarray} 
 \end{mathletters}
This solution expresses the fluctuation function $G_{rrrr}$,
the linear response functions of the $3^{\rm rd}$ order fluctuations
$G_{rrra}$, $G_{rrar}$, $G_{rarr}$, $G_{arrr}$, and 3 of the non-linear
response functions of the $2^{\rm nd}$ order fluctuations ($G_{raar}$, 
$G_{rara}$, $G_{rraa}$) in terms of the fully retarded 4-point vertex 
functions ($3^{\rm rd}$ order non-linear response functions $G_{raaa}$, 
$G_{araa}$, $G_{aara}$, and $G_{aaar}$) and the remaining $2^{\rm nd}$
order response functions of the $2^{\rm nd}$ order fluctuations 
($G_{arra}$, $G_{arar}$, $G_{aarr}$). Although these equations still
have the structure of a fluctuation-dissipation theorem, they are clearly
much more involved than for the simpler cases $n=2,3$ since fluctuation 
and response functions of different order simultaneously.

Similar to Eq.~(\ref{G3vector}) we can express the 4-point Green
function in the single-time representation in terms of tensors
constructed from outer products of 2-component column vectors:
 \begin{eqnarray}
 \label{G4vector}
     &&\Bigl(G_{a_1 a_2 a_3 a_4} \Bigr) = 
          -G_{raaa}{1\choose 1}{n_2\choose {1+n_2}}
                  {n_3\choose {1+n_3}}{n_4\choose {1+n_4}}
     \nonumber\\
     &&-G^*_{raaa}\Bigl[(1+n_2)(1+n_3)(1+n_4)-n_2 n_3 n_4\Bigr]
         {n_1\choose {1+n_1}}{1\choose 1}{1\choose 1}{1\choose 1}
     \nonumber\\
     &&+{1\over 2}G_{araa}{n_1\choose {1+n_1}}{1\choose 1}\Biggl[
                  {n_3\choose {1+n_3}}{{1+n_4}\choose n_4}+
                  {{1+n_3}\choose n_3}{n_4\choose {1+n_4}}\Biggr]
     \nonumber\\
     &&-{1\over 2}G^*_{araa}{{1+n_1}\over n_2}{1\choose 1}{n_2\choose {1+n_2}}
       \Biggl[{{1+n_3}\over n_4}{1\choose 1}{n^2_4\choose {(1+n_4)^2}}+
              {{1+n_4}\over n_3}{n^2_3\choose {(1+n_3)^2}}{1\choose 1}\Biggr]
     \nonumber\\
     &&+{1\over 2}G_{aara}{n_1\choose {1+n_1}}\Biggl[
                  {n_2\choose {1+n_2}}{1\choose 1}{{1+n_4}\choose n_4}+
                  {{1+n_2}\choose n_2}{1\choose 1}{n_4\choose {1+n_4}}\Biggr]
     \nonumber\\
      &&-{1\over 2}G^*_{aara}{{1+n_1}\over n_3}{1\choose 1}
       \Biggl[{{1+n_2}\over n_4}{1\choose 1}{n_3\choose {1+n_3}}
              {n^2_4\choose {(1+n_4)^2}}+
              {{1+n_4}\over n_2}{n^2_2\choose {(1+n_2)^2}}
              {n_3\choose {1+n_3}}{1\choose 1}\Biggr]
     \nonumber\\
     &&+{1\over 2}G_{aaar}{n_1\choose {1+n_1}}\Biggl[
                  {n_2\choose {1+n_2}}{{1+n_3}\choose n_3}+
                  {{1+n_2}\choose n_2}{n_3\choose {1+n_3}}\Biggr]{1\choose 1}
     \nonumber\\
     &&-{1\over 2}G^*_{aaar}{{1+n_1}\over n_4}{1\choose 1}
       \Biggl[{{1+n_2}\over n_3}{1\choose 1}{n^2_3\choose {(1+n_3)^2}}+
              {{1+n_3}\over n_2}{n^2_2\choose {(1+n_2)^2}}{1\choose 1}\Biggr]
              {n_4\choose {1+n_4}}
     \nonumber\\
     &&+{1\over 2}G_{arra}{n_1\choose {1+n_1}}{1\choose 1}{1\choose 1}
                  {n_4\choose {1+n_4}}
       +{1\over 2}G^*_{arra}N^{(14)}_{(23)}{1\choose 1}{n_2\choose {1+n_2}}
        {n_3\choose {1+n_3}}{1\choose 1}
     \nonumber\\
     &&+{1\over 2}G_{arar}{n_1\choose {1+n_1}}{1\choose 1}
                  {n_3\choose {1+n_3}}{1\choose 1}
       +{1\over 2}G^*_{arar}N^{(13)}_{(24)}{1\choose 1}{n_2\choose {1+n_2}}
        {1\choose 1}{n_4\choose {1+n_4}}
     \nonumber\\
     &&+{1\over 2}G_{aarr}{n_1\choose {1+n_1}}
                  {n_2\choose {1+n_2}}{1\choose 1}{1\choose 1}
       +{1\over 2}G^*_{aarr}N^{(12)}_{(34)}{1\choose 1}
        {1\choose 1}{n_3\choose {1+n_3}}{n_4\choose {1+n_4}}\, .
 \end{eqnarray} 
In deriving this expression we have made frequent use of the following
relations which hold in global thermal equilibrium:
 \begin{mathletters}
 \label{distri}
 \begin{eqnarray}
  n(-x) &=& -\Bigl( 1+n(x) \Bigr)\, , 
 \label{distri1}\\
  {{n(x_1) n(x_2) n(x_3)}\over n(x_1+x_2+x_3)} &=&
  \Bigl(1+n(x_1)\Bigr) \Bigl(1+n(x_2)\Bigr) \Bigl(1+n(x_3)\Bigr)
   -n(x_1) n(x_2) n(x_3) \, .
 \label{distri2}
 \end{eqnarray} 
 \end{mathletters}
They are a special case of a general relation proved in the appendix
of Ref.~\cite{HWH98}.

\section{Generalized fluctuation-dissipation theorem for 
         amputated 1PI $\lowercase{n}$-point vertex functions}
\label{sec4}

One can derive in a similar way a nonlinear generalization of the FDT 
for the amputated 1PI $n$-point vertex functions. In analogy to 
(\ref{naeo}) we define 
 \begin{mathletters}
 \label{Gamanaeo}
 \begin{eqnarray}
   \Gamma^{(n)}_{n^e_a,(a_1 \dots a_n)}(1, \dots, n) 
   &=&2^{n\over 2}       
      \sum_{\alpha_1,\dots,\alpha_n=a,r \atop
             n_a(\alpha_1,\dots,\alpha_n)=n^e_a}
      \Gamma^{(n)}_{\alpha_1 \dots \alpha_n}(1, \dots, n)
      Q_{\alpha_1 a_1} \cdots Q_{\alpha_n a_n}\, ,
  \label{Gamanaeo1}\\
   \Gamma^{(n)}_{n^o_a,(a_1 \dots a_n)}(1, \dots, n)
   &=&2^{n\over 2}       
      \sum_{\alpha_1,\dots,\alpha_n=a,r \atop
             n_a(\alpha_1,\dots,\alpha_n)=n^o_a}
      \Gamma^{(n)}_{\alpha_1 \dots \alpha_n}(1, \dots, n)
      Q_{\alpha_1 a_1} \cdots Q_{\alpha_n a_n}\, .
  \label{Gamanaeo2}
 \end{eqnarray} 
 \end{mathletters}
Due to (\ref{Gamarrr}) $n_a^e$ must be $\geq 2$ and $n_a^o \geq 1$.
Using Eqs.~(\ref{Q}) and (\ref{Gamatrans}) we then find instead of 
Eq.~(\ref{sumnaeo}) 
 \begin{mathletters}
 \label{Gamasumnaeo}
 \begin{eqnarray}
  \sum_{2\leq n^e_a\leq n}
       \Gamma^{(n)}_{n^e_a,(a_1 \dots a_n)}(1, \dots, n) &=&
       2^{n-2} \Bigl[ \Gamma^{(n)}_{a_1 \dots a_n}(1, \dots, n) + 
       \Gamma^{(n)}_{{\bar a}_1 \dots {\bar a}_n}(1, \dots, n) 
       \Bigr]\, ,
  \label{Gamasumnaeo1}\\
  \sum_{1\leq n^o_a\leq n}
       \Gamma^{(n)}_{n^o_a,(a_1 \dots a_n)}(1, \dots, n) &=&
       2^{n-2} \Bigl[ \Gamma^{(n)}_{a_1 \dots a_n}(1, \dots, n) - 
       \Gamma^{(n)}_{{\bar a}_1 \dots {\bar a}_n}(1, \dots, n) 
       \Bigr]\, .
  \label{Gamasumnaeo2}
 \end{eqnarray} 
 \end{mathletters}
With the help of the KMS condition (\ref{KMSb}) and the tilde 
conjugation relation (\ref{Gamatildea}) we finally find
 \begin{mathletters}
 \label{GamaFDT}
 \begin{eqnarray}
   && {\rm Im}\left[\sum_{2\leq n^e_a\leq n}
      \Gamma^{(n)}_{n^e_a,(a_1 \dots a_n)}(k_1, \dots, k_n)\right] 
  \nonumber\\
   &&\qquad =\coth\left( {\beta\over 2}\sum_{\{i | a_i=2\}} k^0_i \right)
      {\rm Im}\left[ \sum_{1\leq n^o_a\leq n}
      \Gamma^{(n)}_{n^o_a,(a_1 \dots a_n)}(k_1, \dots, k_n)\right]\, ,
 \label{GamaFDTa}\\
   &&  {\rm Re}\left[\sum_{2\leq n^e_a\leq n}
      \Gamma^{(n)}_{n^e_a,(a_1 \dots a_n)}(k_1, \dots, k_n)\right]
  \nonumber\\
   &&\qquad =\tanh\left( {\beta\over 2}\sum_{\{i | a_i=2\}} k^0_i \right)
      {\rm Re}\left[ \sum_{1\leq n^o_a\leq n}
      \Gamma^{(n)}_{n^o_a,(a_1 \dots a_n)}(k_1, \dots, k_n)\right]\, .
 \label{GamaFDTb}
 \end{eqnarray} 
 \end{mathletters}
These relations are the nonlinear generalization of the FDT for the
amputated 1PI $n$-point vertex function. Contrary to the case for the 
$n$-point Green functions, we here obtain formally the same relations
for even and odd values of $n$. Eqs.~(\ref{GamaFDT}) provide $2^{n-1}$ 
pairs of (real) relations among the (complex) components of 
$\Gamma^{(n)}_{a_1 \dots a_n}$. Together with Eq.~(\ref{Gamarrr}) this
leaves at most $2^{n-1}-1$ independent complex components for
the $n$-point vertex.

\subsection{2-point self-energy}
\label{sec4a}

Making use of Eqs.~(\ref{Gamarrr}) and (\ref{Gamatrans}) in the 
definition (\ref{Gamanaeo}), we find
 \begin{mathletters}
 \label{Ga2naeo}
 \begin{eqnarray}
   \Gamma^{(2)}_{1,(22)} = -\Gamma^{(2)}_{ra}-\Gamma^{(2)}_{ar}\, ,
   \qquad\qquad 
   \Gamma^{(2)}_{2,(22)} &=& \Gamma^{(2)}_{aa}\, ;
  \label{Ga2naeo1}\\
   \Gamma^{(2)}_{1,(21)} = \Gamma^{(2)}_{ra}-\Gamma^{(2)}_{ar}\, ,
   \qquad\qquad\quad 
   \Gamma^{(2)}_{2,(21)} &=& -\Gamma^{(2)}_{aa}\, .
  \label{Ga2naeo2}
 \end{eqnarray} 
 \end{mathletters}
Substitution into (\ref{GamaFDT}) yields
 \begin{mathletters}
 \label{Ga2eq}
 \begin{eqnarray}
   &&{\rm Im}\Bigl[\Gamma^{(2)}_{ra}(k_1, k_2)
                 + \Gamma^{(2)}_{ar}(k_1, k_2)\Bigr] = 0\, , 
  \label{Ga2eq1}\\
   &&{\rm Re\,}\Gamma^{(2)}_{aa}(k_1, k_2) = 0\, ; 
  \label{Ga2eq2}\\
   -&&{\rm Im\,}\Gamma^{(2)}_{aa}(k_1, k_2) = \coth{{\beta k^0_1}\over 2}\,
      {\rm Im}\Bigl[ \Gamma^{(2)}_{ra}(k_1, k_2)
                    -\Gamma^{(2)}_{ar}(k_1, k_2)\Bigr]  \, , 
  \label{Ga2eq3}\\
   -&&{\rm Re\,}\Gamma^{(2)}_{rr}(k_1, k_2) = \tanh{{\beta k^0_1}\over 2}\,
      {\rm Re\,}\Bigl[ \Gamma^{(2)}_{ra}(k_1, k_2)
                      -\Gamma^{(2)}_{ar}(k_1, k_2)\Bigr]  \, . 
  \label{Ga2eq4}
 \end{eqnarray} 
 \end{mathletters}
Replacing in these equations $\Gamma^{(2)}$ by the 2-point Green function
$G$ and interchanging the indices $r$ and $a$ we recover Eq.(\ref{G2eq}). 
The solution of (\ref{Ga2eq}) is thus straightforward:
 \begin{equation}
 \label{Ga2sol}
   \Gamma^{(2)}_{aa}(k) = \coth{{\beta k^0}\over 2}
      \Bigl[ \Gamma^{(2)}_{ar}(k)-\Gamma^{(2)}_{ra}(k)\Bigr]\, ,\qquad 
   \Gamma^{(2)}_{ar}(k) = \Gamma^{(2)*}_{ra}(k)\, . 
 \end{equation} 
This coincides with the results obtained in
Refs.~\cite{Chou,Carrington}. Here $\Gamma^{(2)}_{ar}$ and
$\Gamma^{(2)}_{ra}$ are the retarded and advanced self-energies,
$\Sigma^{\rm ret}$ and $\Sigma^{\rm adv}$, respectively. 

Using the column vector notation one can write the 2-point vertex function
in the single-time representation as 
 \begin{equation}
 \label{Ga2vector}
   \Bigl(\Gamma^{(2)}_{a_1 a_2}(k)\Bigr) = 
   \Gamma^{(2)}_{ar}(k){1\choose -1} {{1+n(k^0)}\choose -n(k^0)}-
          \Gamma^{(2)}_{ra}(k){n(k^0)\choose {-(1+n(k^0))}}{1\choose -1}\, .
 \end{equation} 
This expression exemplifies the general {\bf substitution rule}:
{\em in order to obtain the vertex functions $\Gamma$ from the
corresponding Green functions $G$, one interchanges all $r$ and $a$
indices and changes the sign in the lower components of all the column
vectors.}

\subsection{Amputated 1PI 3-point vertex}
\label{sec4b}

Substituting the quantities
 \begin{mathletters}
 \label{Ga3naeo}
 \begin{eqnarray}
   && \Gamma^{(3)}_{3,(222)} = \Gamma^{(3)}_{3,(211)} = \Gamma^{(3)}_{3,(121)}
    = \Gamma^{(3)}_{3,(112)} = -\Gamma^{(3)}_{aaa}\, ;
  \label{Ga3naeo1}\\
   && \Gamma^{(3)}_{1,(222)} = -\Gamma^{(3)}_{rra}-\Gamma^{(3)}_{rar}
     -\Gamma^{(3)}_{arr}\, ,\qquad
      \Gamma^{(3)}_{2,(222)} = \Gamma^{(3)}_{raa}+\Gamma^{(3)}_{ara}
     +\Gamma^{(3)}_{aar}\, ;
  \label{Ga3naeo2}\\
   && \Gamma^{(3)}_{1,(211)} = \Gamma^{(3)}_{rra}+\Gamma^{(3)}_{rar}
     -\Gamma^{(3)}_{arr}\, ,\qquad
      \Gamma^{(3)}_{2,(211)} = \Gamma^{(3)}_{raa}-\Gamma^{(3)}_{ara}
     -\Gamma^{(3)}_{aar}\, ;
  \label{Ga3naeo3}\\
   && \Gamma^{(3)}_{1,(121)} = \Gamma^{(3)}_{rra}-\Gamma^{(3)}_{rar}
     +\Gamma^{(3)}_{arr}\, ,\qquad
      \Gamma^{(3)}_{2,(121)} = -\Gamma^{(3)}_{raa}+\Gamma^{(3)}_{ara}
     -\Gamma^{(3)}_{aar}\, ;
  \label{Ga3naeo4}\\
   && \Gamma^{(3)}_{1,(112)} = -\Gamma^{(3)}_{rra}+\Gamma^{(3)}_{rar}
     +\Gamma^{(3)}_{arr}\, ,\qquad
      \Gamma^{(3)}_{2,(112)} = -\Gamma^{(3)}_{raa}-\Gamma^{(3)}_{ara}
     +\Gamma^{(3)}_{aar}
  \label{Ga3naeo5}
 \end{eqnarray} 
 \end{mathletters}
into Eq.~(\ref{GamaFDT}) we find
 \begin{mathletters}
 \label{Ga3eq}
 \begin{eqnarray}
   &&{\rm Im}\Bigl[ \Gamma^{(3)}_{rra}+\Gamma^{(3)}_{rar}+\Gamma^{(3)}_{arr}
     +\Gamma^{(3)}_{aaa}\Bigr] = 0\, , 
  \label{Ga3eq1}\\
   &&{\rm Re}\Bigl[ \Gamma^{(3)}_{raa}+\Gamma^{(3)}_{ara}+\Gamma^{(3)}_{aar}
     \Bigr] = 0\, ; 
  \label{Ga3eq2}\\
   &&{\rm Im}\Bigl[ \Gamma^{(3)}_{raa}-\Gamma^{(3)}_{ara}-\Gamma^{(3)}_{aar}
     \Bigr] = \coth{{\beta k^0_1}\over 2}
     {\rm Im}\Bigl[ \Gamma^{(3)}_{rra}+\Gamma^{(3)}_{rar}-\Gamma^{(3)}_{arr}
     -\Gamma^{(3)}_{aaa}\Bigr]\, ,
  \label{Ga3eq3}\\
   &&{\rm Re}\Bigl[ \Gamma^{(3)}_{raa}-\Gamma^{(3)}_{ara}-\Gamma^{(3)}_{aar}
     \Bigr] = \tanh{{\beta k^0_1}\over 2}
     {\rm Re}\Bigl[ \Gamma^{(3)}_{rra}+\Gamma^{(3)}_{rar}-\Gamma^{(3)}_{arr}
     -\Gamma^{(3)}_{aaa}\Bigr]\, ;
  \label{Ga3eq4}\\
   &&{\rm Im}\Bigl[ -\Gamma^{(3)}_{raa}+\Gamma^{(3)}_{ara}-\Gamma^{(3)}_{aar}
     \Bigr] = \coth{{\beta k^0_2}\over 2}
     {\rm Im}\Bigl[ \Gamma^{(3)}_{rra}-\Gamma^{(3)}_{rar}+\Gamma^{(3)}_{arr}
     -\Gamma^{(3)}_{aaa}\Bigr]\, ,
  \label{Ga3eq5}\\
   &&{\rm Re}\Bigl[ -\Gamma^{(3)}_{raa}+\Gamma^{(3)}_{ara}-\Gamma^{(3)}_{aar}
     \Bigr] = \tanh{{\beta k^0_2}\over 2}
     {\rm Re}\Bigl[ \Gamma^{(3)}_{rra}-\Gamma^{(3)}_{rar}+\Gamma^{(3)}_{arr}
     -\Gamma^{(3)}_{aaa}\Bigr]\, ;
  \label{Ga3eq6}\\
   &&{\rm Im}\Bigl[ -\Gamma^{(3)}_{raa}-\Gamma^{(3)}_{ara}+\Gamma^{(3)}_{aar}
     \Bigr] = \coth{{\beta k^0_3}\over 2}
     {\rm Im}\Bigl[ -\Gamma^{(3)}_{rra}+\Gamma^{(3)}_{rar}+\Gamma^{(3)}_{arr}
     -\Gamma^{(3)}_{aaa}\Bigr]\, ,
  \label{Ga3eq7}\\
   &&{\rm Re}\Bigl[ -\Gamma^{(3)}_{raa}-\Gamma^{(3)}_{ara}+\Gamma^{(3)}_{aar}
     \Bigr] = \tanh{{\beta k^0_3}\over 2}
     {\rm Re}\Bigl[ -\Gamma^{(3)}_{rra}+\Gamma^{(3)}_{rar}+\Gamma^{(3)}_{arr}
     -\Gamma^{(3)}_{aaa}\Bigr]\, .
  \label{Ga3eq8}
 \end{eqnarray} 
 \end{mathletters}
Replacing $\Gamma^{(3)}$ by the 3-point Green function $G$ and 
interchanging the indices $r$ and $a$ we recover Eq.~(\ref{G3eq}). 
Choosing as independent components the fully retarded vertices
$\Gamma^{(3)}_{arr}$, $\Gamma^{(3)}_{rar}$ and $\Gamma^{(3)}_{rra}$, 
we thus obtain directly the solution 
 \begin{mathletters}
 \label{Ga3sol}
 \begin{eqnarray}
   \Gamma^{(3)}_{ara} &=& N_1 ( \Gamma^{(3)*}_{rar}-\Gamma^{(3)}_{rra} )+
               N_3 ( \Gamma^{(3)*}_{rar}-\Gamma^{(3)}_{arr} )\, , 
  \label{Ga3sol1}\\
   \Gamma^{(3)}_{raa} &=& N_2 ( \Gamma^{(3)*}_{arr}-\Gamma^{(3)}_{rra} )+
               N_3 ( \Gamma^{(3)*}_{arr}-\Gamma^{(3)}_{rar} )\, , 
  \label{Ga3sol2}\\
   \Gamma^{(3)}_{aar} &=& N_1 ( \Gamma^{(3)*}_{rra}-\Gamma^{(3)}_{rar} )+
               N_2 ( \Gamma^{(3)*}_{rra}-\Gamma^{(3)}_{arr} )\, , 
  \label{Ga3sol3}\\
   \Gamma^{(3)}_{aaa} &=& \Gamma^{(3)*}_{arr}+ \Gamma^{(3)*}_{rar}
               +\Gamma^{(3)*}_{rra} +N_2 N_3 ( \Gamma^{(3)}_{arr}
               +\Gamma^{(3)*}_{arr} )
  \nonumber\\
           & &+N_1 N_3 ( \Gamma^{(3)}_{rar}+\Gamma^{(3)*}_{rar} ) 
              +N_1 N_2 ( \Gamma^{(3)}_{rra}+\Gamma^{(3)*}_{rra} )\, .
  \label{Ga3sol4}
 \end{eqnarray} 
 \end{mathletters}
These relations are the nonlinear generalization of the FDT for the 
amputated 1PI 3-point vertex function. 

In terms of outer products of column vectors the 3-point vertex
function in the single-time representation can be expressed as 
 \begin{eqnarray}
 \label{Ga3vector}
   \Bigl(\Gamma^{(3)}_{a_1 a_2 a_3}\Bigr) &=&
       \Gamma^{(3)}_{arr}{1\choose -1}{n_2\choose {-(1+n_2)}}
                  {n_3\choose {-(1+n_3)}}
     \nonumber\\
   &-& \Gamma^{(3)*}_{arr}\Bigl[(1+n_2)(1+n_3)-n_2 n_3\Bigr]
         {n_1\choose {-(1+n_1)}}{1\choose -1}{1\choose -1}
     \nonumber\\
   &+& \Gamma^{(3)}_{rar}{n_1\choose {-(1+n_1)}}{1\choose -1}
         {n_3\choose {-(1+n_3)}}
     \nonumber\\
   &-& \Gamma^{(3)*}_{rar}\Bigl[(1+n_1)(1+n_3)-n_1 n_3\Bigr]
         {1\choose -1}{n_2\choose {-(1+n_2)}}{1\choose -1}
     \nonumber\\
   &+& \Gamma^{(3)}_{rra}{n_1\choose {-(1+n_1)}}{n_2\choose {-(1+n_2)}}
         {1\choose -1}
     \nonumber\\
   &-& \Gamma^{(3)*}_{rra}\Bigl[(1+n_1)(1+n_2)-n_1 n_2\Bigr]
         {1\choose -1}{1\choose -1}{n_3\choose {-(1+n_3)}}\, ,
 \end{eqnarray} 
in accordance with the above substitution rule (cf. Eq.~(\ref{G3vector})).

\subsection{Amputated 1PI 4-point vertex}
\label{sec4c}

Repeating the same technical steps for the 4-point function we find
 \begin{mathletters}
 \label{Ga4sol}
 \begin{eqnarray}
   \Gamma^{(4)}_{aaaa} =&& -N_2 N_3 N_4 \Gamma^{(4)}_{arrr}
    +\Bigl( N_2 N_3 N_4 +N_2 +N_3 +N_4\Bigr) \Gamma^{(4)*}_{arrr}
    +N_1 N_3 N_4 \Gamma^{(4)}_{rarr} 
   \nonumber\\
    &&+N_2 \Bigl( N^{(14)}_{(23)}
      +N_4^2 N^{(13)}_{(24)}\Bigr) \Gamma^{(4)*}_{rarr}
      +N_1 N_2 N_4 \Gamma^{(4)}_{rrar} +N_3 \Bigl( N^{(12)}_{(34)}
      +N_2^2 N^{(14)}_{(23)}\Bigr) \Gamma^{(4)*}_{rrar}
   \nonumber\\
    &&+N_1 N_2 N_3 \Gamma^{(4)}_{rrra} +N_4 \Bigl( N^{(13)}_{(24)}
      +N_3^2 N^{(12)}_{(34)}\Bigr) \Gamma^{(4)*}_{rrra}
      +N_1 N_4 \Gamma^{(4)}_{raar}+N_2 N_3 N^{(14)}_{(23)}\Gamma^{(4)*}_{raar} 
   \nonumber\\
    &&+N_1 N_3 \Gamma^{(4)}_{rara}
      +N_2 N_4 N^{(13)}_{(24)}\Gamma^{(4)*}_{rara}+N_1 N_2 \Gamma^{(4)}_{rraa}
      +N_3 N_4 N^{(12)}_{(34)}\Gamma^{(4)*}_{rraa}\, ,
  \label{Ga4sol1}\\
   \Gamma^{(4)}_{aaar} =&&N_2 N_3\Gamma^{(4)}_{arrr}
      -N_2 N_4 N^{(13)}_{(24)}\Gamma^{(4)*}_{rarr} 
      -N_3 N_4 N^{(12)}_{(34)}\Gamma^{(4)*}_{rrar}
      -\Bigl( N^{(13)}_{(24)}+N_3^2 N^{(12)}_{(34)}\Bigr) \Gamma^{(4)*}_{rrra}
   \nonumber\\
    &&-N_1 \Gamma^{(4)}_{raar}-N_2 N^{(13)}_{(24)}\Gamma^{(4)*}_{rara}
      -N_3 N^{(12)}_{(34)}\Gamma^{(4)*}_{rraa}\, ,
  \label{Ga4sol2}\\
   \Gamma^{(4)}_{aara} =&&N_2 N_4\Gamma^{(4)}_{arrr}
      -N_2 N_3 N^{(14)}_{(23)}\Gamma^{(4)*}_{rarr} 
      -\Bigl( N^{(12)}_{(34)}+N_2^2 N^{(14)}_{(23)}\Bigr) \Gamma^{(4)*}_{rrar}
      -N_3 N_4 N^{(12)}_{(34)}\Gamma^{(4)*}_{rrra}
   \nonumber\\
    &&-N_2 N^{(14)}_{(23)}\Gamma^{(4)*}_{raar}-N_1 \Gamma^{(4)}_{rara}
      -N_4 N^{(12)}_{(34)}\Gamma^{(4)*}_{rraa}\, ,
  \label{Ga4sol3}\\
   \Gamma^{(4)}_{araa} =&&N_3 N_4\Gamma^{(4)}_{arrr} 
      -\Bigl( N^{(14)}_{(23)}+N_4^2 N^{(13)}_{(24)}\Bigr) \Gamma^{(4)*}_{rarr}
      -N_2 N_3 N^{(14)}_{(23)}\Gamma^{(4)*}_{rrar}
      -N_2 N_4 N^{(13)}_{(24)}\Gamma^{(4)*}_{rrra}
   \nonumber\\
    &&-N_3 N^{(14)}_{(23)}\Gamma^{(4)*}_{raar}
      -N_4 N^{(13)}_{(24)}\Gamma^{(4)*}_{rara}-N_1 \Gamma^{(4)}_{rraa}\, ,
  \label{Ga4sol4}\\
   \Gamma^{(4)}_{raaa} =&&\Bigl( 1+ N_2 N_3+ N_2 N_4
      +N_3 N_4\Bigr)\Gamma^{(4)*}_{arrr} 
      -N_3 N_4 \Gamma^{(4)}_{rarr}-N_2 N_4 \Gamma^{(4)}_{rrar}
      -N_2 N_3 \Gamma^{(4)}_{rrra}
   \nonumber\\
    &&-N_4 \Gamma^{(4)}_{raar}-N_3 \Gamma^{(4)}_{rara}-N_2 \Gamma^{(4)}_{rraa}
  \label{Ga4sol5}\\
   \Gamma^{(4)}_{arra} =&& -N_4 \Gamma^{(4)}_{arrr}
      +N_3 N^{(14)}_{(23)}\Gamma^{(4)*}_{rarr}
      +N_2 N^{(14)}_{(23)}\Gamma^{(4)*}_{rrar}-N_1 \Gamma^{(4)}_{rrra}
      +N^{(14)}_{(23)}\Gamma^{(4)*}_{raar}\, , 
  \label{Ga4sol6}\\
   \Gamma^{(4)}_{arar} =&& -N_3 \Gamma^{(4)}_{arrr}
      +N_4 N^{(13)}_{(24)}\Gamma^{(4)*}_{rarr}
      -N_1 \Gamma^{(4)}_{rrar}+N_2 N^{(13)}_{(24)}\Gamma^{(4)*}_{rrra}
      +N^{(13)}_{(24)}\Gamma^{(4)*}_{rara}\, , 
  \label{Ga4sol7}\\
   \Gamma^{(4)}_{aarr} =&& -N_2 \Gamma^{(4)}_{arrr}-N_1 \Gamma^{(4)}_{rarr}
      +N_4 N^{(12)}_{(34)}\Gamma^{(4)*}_{rrar}
      +N_3 N^{(12)}_{(34)}\Gamma^{(4)*}_{rrra}
      +N^{(12)}_{(34)}\Gamma^{(4)*}_{rraa}\, . 
  \label{Ga4sol8}
 \end{eqnarray} 
 \end{mathletters}
These relations are the nonlinear generalization of the FDT for the
amputated 1PI 4-point vertex function. They can again be obtained by 
substituting in (\ref{G4sol}) $G$ by $\Gamma^{(4)}$ and interchanging 
$r$ with $a$. The corresponding column vector representation is
obtained by applying the above-mentioned substitution rule to
Eq.~(\ref{G4vector}). 

\section{Conclusions}
\label{sec5}

Using the transformation between the single-time and physical 
representations of real-time thermal $n$-point functions, the
tilde conjugation relation and the KMS condition, we have derived
the nonlinear generalization of the fluctuation-dissipation theorem
for real-time $n$-point Green functions and 1PI amputated vertex 
functions at finite temperature. This generalized FDT is formally
identical for disconnected and connected $n$-point Green functions.
The FDT for the amputated 1PI vertex functions is obtained from that
for the Green functions by interchanging the indices $r$ and $a$. 

The generalized FDT for nonlinear response functions provides
model-independent relations between the various components
of the RTF thermal $n$-point functions which can be used as consistency 
checks for approximations. The results derived in this paper can be used
for elementary or composite bosonic fields with arbitrary interactions.

For the cases $n=2$ and $3$ the results from the generalized FDT 
were shown to reproduce known relationships. The new results for 
the case $n=4$ are expected to be useful for the derivation of spectral 
representations for the 4-point function and in a calculation
of transport coefficients like shear viscosity in scalar field theories.
An experimental verification of these relations may be possible
in applications to the recent developments of picosecond pulse 
techniques and multichannel data acquisition in nonlinear systems 
\cite{Halley}.

\acknowledgments
E.W. gratefully acknowledges support by the Alexander von Humboldt
Foundation through a Research Fellowship. The work of U.H. was 
supported in part by DFG, BMBF, and GSI. He wishes to thank the Institute 
for Nuclear Theory at the University of Washington in Seattle, where
this work was written, for kind hospitality.

\appendix
\section{Relation between $\lowercase{r/a}$- and $R/A$-functions}
\label{appa}

In this appendix we derive the relation between the $r/a$ vertex
functions used in the text and the $R/A$-functions introduced by
Aurenche and Becherrawy in Ref.~\cite{Aurenche}. As shown in
\cite{Aurenche}, the propagator $G$ in the single-time representation
can be diagonalized with two matrices $U$ and $V$,
 \begin{equation}
 \label{Ghat}
   G = U {\widehat G } V\, ,
 \end{equation} 
where 
 \begin{equation}
 \label{GAB}
   {\widehat G} = \left( \begin{array}{cc}
                      G_{RR}, & G_{RA}\\
                      G_{AR}, & G_{AA}
                     \end{array} \right)
            = \left( \begin{array}{cc}
                      G^{\rm ret}, & 0\\
                      0, & G^{\rm adv} 
                     \end{array} \right)\, ,
 \end{equation} 
and we choose
 \begin{mathletters}
 \label{UV}
 \begin{eqnarray}
   U = \left( \begin{array}{cc}
                      U_{1R}, & U_{1A}\\
                      U_{2R}, & U_{2A}
                     \end{array} \right)
            = \left( \begin{array}{cc}
                      1, & n\\
                      1, & 1+n 
                     \end{array} \right)\, ,
 \label{UV1}\\
   V = \left( \begin{array}{cc}
                      V_{R1}, & V_{R2}\\
                      V_{A1}, & V_{A2}
                     \end{array} \right)
            = \left( \begin{array}{cc}
                      1+n, & n\\
                      -1, & -1 
                     \end{array} \right)
 \label{UV2}
 \end{eqnarray} 
 \end{mathletters}
as in Ref.\cite{Henning}. 
 
    In the CTP formalism the relation between propagator
$G$ and the advanced, retarded and correlation functions is\cite{Chou}
 \begin{equation}
 \label{Gbar}
   G = Q^{\dagger} {\bar G } Q\, ,
 \end{equation} 
where 
 \begin{mathletters}
 \label{QQ}
 \begin{eqnarray}
   {\bar G} &=& \left( \begin{array}{cc}
                      G_{aa}, & G_{ar}\\
                      G_{ra}, & G_{rr}
                     \end{array} \right)
            = \left( \begin{array}{cc}
                      0, & G^{\rm adv}\\
                      G^{\rm ret}, & G^{\rm cor} 
                     \end{array} \right)\, ,
 \label{QQ1}\\
   Q &=& \left( \begin{array}{cc}
                      Q_{a1}, & Q_{a2}\\
                      Q_{r1}, & Q_{r2}
                     \end{array} \right)
            = {1\over \sqrt{2}} \left( \begin{array}{cc}
                      1, & -1\\
                      1, & 1 
                     \end{array} \right)\, ,
 \label{QQ2}\\
   Q^{\dagger}=Q^{-1} &=& \left( \begin{array}{cc}
                      Q^{\dagger}_{1a}, & Q^{\dagger}_{1r}\\
                      Q^{\dagger}_{2a}, & Q^{\dagger}_{2r}
                     \end{array} \right)
            = {1\over \sqrt{2}} \left( \begin{array}{cc}
                      1, & 1\\
                      -1, & 1 
                     \end{array} \right)\, .
 \label{QQ3}
 \end{eqnarray} 
 \end{mathletters}

Combining Eqs.~(\ref{Ghat}) and (\ref{Gbar}) leads to 
 \begin{equation}
 \label{Grela}
   {\widehat G } = U^{-1} Q^{\dagger} {\bar G} Q V^{-1}\, .
 \end{equation} 
 
\subsection{Self energy function}
\label{appa1}

Substituting the Schwinger-Dyson equation
 \begin{mathletters}
 \label{dyson}
 \begin{eqnarray}
   {\widehat G }^{-1} &=& {\widehat G}_0^{-1} - {\widehat \Sigma}\, ,
 \label{dyson1}\\
   {\bar G }^{-1} &=& {\bar G}_0^{-1} - {\bar \Sigma}
 \label{dyson2}
 \end{eqnarray} 
 \end{mathletters}
into Eq.~(\ref{Grela}) we deduce
 \begin{equation}
 \label{sigmarela}
   {\widehat \Sigma } = V Q^{\dagger} {\bar \Sigma} Q U\, ,
 \end{equation} 
where
 \begin{equation}
 {\widehat \Sigma} = \left( \begin{array}{cc}
                      \Sigma_{RR}, & \Sigma_{RA}\\
                      \Sigma_{AR}, & \Sigma_{AA} 
                     \end{array} \right)\, , 
 \qquad
 {\bar \Sigma} = \left( \begin{array}{cc}
                      \Gamma^{(2)}_{aa}, & \Gamma^{(2)}_{ar}\\
                      \Gamma^{(2)}_{ra}, & \Gamma^{(2)}_{rr} 
                     \end{array} \right)\, .
 \label{sigma}
 \end{equation} 
The explicit matrix form of Eq.~(\ref{sigmarela}) is
 \begin{eqnarray}
 \label{sigmamatrix}
   &&\qquad\qquad\qquad\qquad\qquad\qquad
     \left( \begin{array}{cc}
          \Sigma_{RR}, & \Sigma_{RA}\\
          \Sigma_{AR}, & \Sigma_{AA} 
          \end{array} \right) =
  \nonumber\\ 
   &&\left( \begin{array}{cc}
          \Gamma^{(2)}_{ar} + (1+2 n)\Gamma^{(2)}_{rr}, 
        & -{1\over 2}\Gamma^{(2)}_{aa} + {1\over 2} (1+2 n) 
          (\Gamma^{(2)}_{ar}-\Gamma^{(2)}_{ra}) 
          +{1\over 2} (1+2 n)^2 \Gamma^{(2)}_{rr}\\
          -2\Gamma^{(2)}_{rr}, 
        & \Gamma^{(2)}_{ra} - (1+2 n)\Gamma^{(2)}_{rr} 
          \end{array} \right)\, .
 \end{eqnarray} 
Inserting the FDTs (\ref{Ga2sol}) and Eq.~(\ref{Gamarrr}) into the
above expression we get following relations:
 \begin{equation}
 \label{sigmarel}
  \Sigma_{RR} = \Gamma^{(2)}_{ar},\quad
  \Sigma_{AA} = \Gamma^{(2)}_{ra},\quad
  \Sigma_{RA} = \Sigma_{AR} = 0,\quad
  \Sigma_{RR} = \Sigma^*_{AA} .
 \end{equation}
The first two of these equations agree with Eqs.~(4.3) and (4.7)
in Ref.~\cite{Aurenche}. 

Please note that $\Sigma_{RR}$ and $\Sigma_{AA}$ do not vanish, 
in contrast to the multi-point cases ({\em c.f.}
Eqs.~(\ref{Gamarela1}) and (\ref{hatgama41})).

\subsection{n-point vertex function ($n\ge 3$)}
\label{appa2}

As shown in Ref.~\cite{Aurenche}, the thermal distribution functions 
which occur in the finite temperature propagators via the matrices 
$U(k^0)$ and $V(k^0)$ can be fully absorbed into the vertex
functions. To this end one combines the matrices $U$ and $V$ in the
propagators (\ref{Ghat}) with the vertex to which $G$ attaches;
$U(k^0)$ is associated with an outgoing line while $V(k^0)$ is
associated to an incoming line with momentum $k$. This defines the
so-called $R/A$ vertex functions~\cite{Aurenche}. Thus the $n$-point
$R/A$ vertex function with all incoming momenta $k_1,k_2,\dots,k_n$
can be expressed as 
 \begin{equation}
 \label{Gamalambda}
   {\widehat \Gamma}^{(n)}_{\Lambda_1\dots\Lambda_n}(k_1\dots k_n) = 
     V_{\Lambda_1 a_1}(k^0_1)\dots V_{\Lambda_n a_n}(k^0_n)
     \Gamma^{(n)}_{a_1\dots a_n}(k_1\dots k_n)\, ,
 \end{equation} 
where $\Lambda_1, \dots, \Lambda_n=R,A$. 
${\widehat \Gamma}^{(n)}_{\Lambda_1\dots\Lambda_n}$ incorporates all 
temperature dependence of the finite-temperature retarded and advanced
fields\cite{Weldon}. Inserting Eq.~(\ref{Gamatrans}) into the above
equation and defining 
 \begin{equation}
 \label{P}
  P_{\Lambda_i \alpha_i}(k^0_i)=
    \sqrt{2}\, V_{\Lambda_i a_j}(k^0_i) Q^{\dagger}_{a_j \alpha_i} \,,
 \end{equation} 
we obtain the relation between the $r/a$-functions used in the text
and the $R/A$-functions of Aurenche and Becherrawy~\cite{Aurenche}:
 \begin{equation}
 \label{RArela}
   {\widehat \Gamma}^{(n)}_{\Lambda_1\dots\Lambda_n}(k_1\dots k_n) =2^{1-n} 
     P_{\Lambda_1 \alpha_1}(k^0_1)\dots P_{\Lambda_n \alpha_n}(k^0_n)
     \Gamma^{(n)}_{\alpha_1\dots \alpha_n}(k_1\dots k_n)\, ,
 \end{equation} 
where $\alpha_1, \dots, \alpha_n=r,a$. The explicit form of
Eq.~(\ref{P}) reads
 \begin{eqnarray}
 \label{Pcompon}
     P_{Ra}(k^0_i) &=& 1\, ,\qquad P_{Rr}(k^0_i) = N(k^0_i) =1 + 2n(k^0_i)\, ,
   \nonumber\\
     P_{Aa}(k^0_i) &=& 0\, ,\qquad P_{Ar}(k^0_i) = -2\, .
 \end{eqnarray} 
Clearly the $R/A$ vertex functions ${\widehat \Gamma}^{(n)}_{\Lambda_1
  \dots \Lambda_n}$ are linear combinations of the $r/a$ vertex
functions $\Gamma^{(n)}_{\alpha_1\dots \alpha_n}$. We will now show
that the $R/A$ vertex functions have simple relations with
the independent components of the $r/a$ vertex functions after
inserting the generalized FDTs.

\subsubsection{3-point vertex function}
\label{appa2a}

Substituting the FDTs (\ref{Ga3sol}) and Eq.~(\ref{Gamarrr}) into 
Eq.~(\ref{RArela}) for $n$=3 we deduce
 \begin{mathletters}
 \label{Gamarela}
 \begin{eqnarray}
  {\widehat \Gamma}^{(3)}_{AAA} &=& {\widehat \Gamma}^{(3)}_{RRR}=0\, , 
  \label{Gamarela1}\\
  {\widehat \Gamma}^{(3)}_{RAA} &=& \Gamma^{(3)}_{arr}\, , 
  \label{Gamarela2}\\
  {\widehat \Gamma}^{(3)}_{ARA} &=& \Gamma^{(3)}_{rar}\, , 
  \label{Gamarela3}\\
  {\widehat \Gamma}^{(3)}_{AAR} &=& \Gamma^{(3)}_{rra}\, ,
  \label{Gamarela4}\\
  {\widehat \Gamma}^{(3)}_{ARR} &=&
           -{1\over 2} (N_2+N_3) \Gamma^{(3)*}_{arr}\, , 
  \label{Gamarela5}\\
  {\widehat \Gamma}^{(3)}_{RAR} &=&
           -{1\over 2} (N_1+N_3) \Gamma^{(3)*}_{rar}\, , 
  \label{Gamarela6}\\
  {\widehat \Gamma}^{(3)}_{RRA} &=& 
           -{1\over 2} (N_1+N_2) \Gamma^{(3)*}_{rra}\, . 
  \label{Gamarela7}
 \end{eqnarray} 
 \end{mathletters}
These relations can be also obtained in the following simpler
manner. Noticing 
 \begin{equation}
 \label{product}
   V(k^0_i){1\choose -1}={1\choose 0}\, ,\qquad
   V(k^0_i){n(k^0_i)\choose -(1+n(k^0_i))}={0\choose 1}\, ,
 \end{equation} 
and substituting Eq.~(\ref{Ga3vector}) into Eq.~(\ref{Gamalambda}) we have
 \begin{eqnarray}
 \label{hatGa3vector}
   \Bigl({\widehat \Gamma}^{(3)}_{\Lambda_1\Lambda_2\Lambda_3} \Bigr)
   &=&\Gamma^{(3)}_{arr}{1\choose 0}{0\choose 1}
                  {0\choose 1}
    - {1\over 2} (N_2+N_3)\Gamma^{(3)*}_{arr}
         {0\choose 1}{1\choose 0}{1\choose 0}
     \nonumber\\
   &+& \Gamma^{(3)}_{rar}{0\choose 1}{1\choose 0}{0\choose 1}
    - {1\over 2} (N_1+N_3)\Gamma^{(3)*}_{rar}
         {1\choose 0}{0\choose 1}{1\choose 0}
     \nonumber\\
   &+& \Gamma^{(3)}_{rra}{0\choose 1}{0\choose 1}{1\choose 0}
    - {1\over 2} (N_1+N_2)\Gamma^{(3)*}_{rra}
         {1\choose 0}{1\choose 0}{0\choose 1}\, .
 \end{eqnarray} 
Eqs.~(\ref{Gamarela}) can now be directly read off. It should be noted
that the FDTs play an important role in getting Eqs.~(\ref{Gamarela}).

Using ${\widehat \Gamma}^{(3)}_{RAA}$, ${\widehat \Gamma}^{(3)}_{ARA}$ and 
${\widehat \Gamma}^{(3)}_{AAR}$ as independent components,
Eqs.~(\ref{Gamarela}) can be used to rewrite Eq.~(\ref{Ga3vector}) as
 \begin{eqnarray}
 \label{Ga3vect}
   \Bigl(\Gamma^{(3)}_{a_1 a_2 a_3}\Bigr) &=&
       {\widehat \Gamma}^{(3)}_{RAA}{1\choose -1}{n_2\choose {-(1+n_2)}}
                  {n_3\choose {-(1+n_3)}}
     \nonumber\\
   &-& {\widehat \Gamma}^{(3)*}_{RAA}
         {{(1+n_2)(1+n_3)-n_2 n_3}\over {(1+n_1)-n_1}}
         {n_1\choose {-(1+n_1)}}{1\choose -1}{1\choose -1}
     \nonumber\\
   &+& {\widehat \Gamma}^{(3)}_{ARA}{n_1\choose {-(1+n_1)}}{1\choose -1}
         {n_3\choose {-(1+n_3)}}
     \nonumber\\
   &-& {\widehat \Gamma}^{(3)*}_{ARA}
         {{(1+n_1)(1+n_3)-n_1 n_3}\over {(1+n_2)-n_2}}
         {1\choose -1}{n_2\choose {-(1+n_2)}}{1\choose -1}
     \nonumber\\
   &+& {\widehat \Gamma}^{(3)}_{AAR}{n_1\choose {-(1+n_1)}}
         {n_2\choose {-(1+n_2)}}{1\choose -1}
     \nonumber\\
   &-& {\widehat \Gamma}^{(3)*}_{AAR}
         {{(1+n_1)(1+n_2)-n_1 n_2}\over {(1+n_3)-n_3}}
         {1\choose -1}{1\choose -1}{n_3\choose {-(1+n_3)}}\, .
 \end{eqnarray} 

\subsubsection{4-point vertex function}
\label{appa2b}

Similar to the 3-point vertex function, using Eqs.~(\ref{Gamalambda}) or
(\ref{RArela}) we deduce 
 \begin{mathletters}
 \label{hatgama4}
 \begin{eqnarray}
  {\widehat \Gamma}^{(4)}_{AAAA} &=& {\widehat \Gamma}^{(4)}_{RRRR}=0\, , 
  \label{hatgama41}\\
  {\widehat \Gamma}^{(4)}_{RRRA} &=& 
           {{(1+n_1)(1+n_2)(1+n_3)}\over n_4} \Gamma^{(4)*}_{rrra}\, , 
  \label{hatgama42}\\
  {\widehat \Gamma}^{(4)}_{RRAR} &=& 
           {{(1+n_1)(1+n_2)(1+n_4)}\over n_3} \Gamma^{(4)*}_{rrar}\, , 
  \label{hatgama43}\\
  {\widehat \Gamma}^{(4)}_{RARR} &=& 
           {{(1+n_1)(1+n_3)(1+n_4)}\over n_2} \Gamma^{(4)*}_{rarr}\, , 
  \label{hatgama44}\\
  {\widehat \Gamma}^{(4)}_{ARRR} &=& 
           {{(1+n_2)(1+n_3)(1+n_4)}\over n_1} \Gamma^{(4)*}_{arrr}\, , 
  \label{hatgama45}\\
  {\widehat \Gamma}^{(4)}_{RRAA} &=& 
           {1\over 2} N^{(12)}_{(34)}\Bigl( \Gamma^{(4)*}_{rraa}
           +N_3\Gamma^{(4)*}_{rrra}+N_4\Gamma^{(4)*}_{rrar}\Bigr)\, , 
  \label{hatgama46}\\
  {\widehat \Gamma}^{(4)}_{RARA} &=& 
           {1\over 2} N^{(13)}_{(24)}\Bigl( \Gamma^{(4)*}_{rara}
           +N_2\Gamma^{(4)*}_{rrra}+N_4\Gamma^{(4)*}_{rarr}\Bigr)\, , 
  \label{hatgama47}\\
  {\widehat \Gamma}^{(4)}_{RAAR} &=& 
           {1\over 2} N^{(14)}_{(23)}\Bigl( \Gamma^{(4)*}_{raar}
           +N_2\Gamma^{(4)*}_{rrar}+N_3\Gamma^{(4)*}_{rarr}\Bigr)\, , 
  \label{hatgama48}\\
  {\widehat \Gamma}^{(4)}_{AARR} &=& 
           {1\over 2} \Bigl( \Gamma^{(4)}_{rraa}
           +N_3\Gamma^{(4)}_{rrra}+N_4\Gamma^{(4)}_{rrar}\Bigr)\, , 
  \label{hatgama49}\\
  {\widehat \Gamma}^{(4)}_{ARAR} &=& 
           {1\over 2} \Bigl( \Gamma^{(4)}_{rara}
           +N_2\Gamma^{(4)}_{rrra}+N_4\Gamma^{(4)}_{rarr}\Bigr)\, , 
  \label{hatgama410}\\
  {\widehat \Gamma}^{(4)}_{ARRA} &=& 
           {1\over 2} \Bigl( \Gamma^{(4)}_{raar}
           +N_2\Gamma^{(4)}_{rrar}+N_3\Gamma^{(4)}_{rarr}\Bigr)\, , 
  \label{hatgama411}\\
  {\widehat \Gamma}^{(4)}_{AAAR} &=& -\Gamma^{(4)}_{rrra}\, , 
  \label{hatgama412}\\
  {\widehat \Gamma}^{(4)}_{AARA} &=& -\Gamma^{(4)}_{rrar}\, , 
  \label{hatgama413}\\
  {\widehat \Gamma}^{(4)}_{ARAA} &=& -\Gamma^{(4)}_{rarr}\, , 
  \label{hatgama414}\\
  {\widehat \Gamma}^{(4)}_{RAAA} &=& -\Gamma^{(4)}_{arrr} 
  \label{hatgama415}
 \end{eqnarray} 
 \end{mathletters}
for 4-point vertex function.

Using ${\widehat \Gamma}^{(4)}_{RAAA}$, ${\widehat \Gamma}^{(4)}_{ARAA}$,
${\widehat \Gamma}^{(4)}_{AARA}$, ${\widehat \Gamma}^{(4)}_{AAAR}$,
${\widehat \Gamma}^{(4)}_{RAAR}$, ${\widehat \Gamma}^{(4)}_{RARA}$,
${\widehat \Gamma}^{(4)}_{RRAA}$ as independent components, 
similar to Eq.~(\ref{Ga3vect}) we can express the 4-point vertex
function in the single-time representation in terms of column
vectors as \cite{HWH98}
 \begin{eqnarray}
 \label{Ga4vect}
     \Gamma^{(4)}_{a_1 a_2 a_3 a_4}  
     &=&{\widehat \Gamma}^{(4)}_{RAAA}{1\choose -1}{n_2\choose {-(1+n_2)}}
         {n_3\choose {-(1+n_3)}}{n_4\choose {-(1+n_4)}}
     \nonumber\\
     &+&{\widehat \Gamma}^{(4)*}_{RAAA}{{(1+n_2)(1+n_3)(1+n_4)
         -n_2 n_3 n_4}\over {(1+n_1)-n_1}}
         {n_1\choose {-(1+n_1)}}{1\choose -1}{1\choose -1}{1\choose -1}
     \nonumber\\
     &+&{\widehat \Gamma}^{(4)}_{ARAA}{n_1\choose {-(1+n_1)}}{1\choose -1}
         {n_3\choose {-(1+n_3)}}{n_4\choose {-(1+n_4)}}
     \nonumber\\
     &+&{\widehat \Gamma}^{(4)*}_{ARAA}{{(1+n_1)(1+n_3)(1+n_4)
         -n_1 n_3 n_4}\over {(1+n_2)-n_2}}
         {1\choose -1}{n_2\choose {-(1+n_2)}}{1\choose -1}{1\choose -1}
     \nonumber\\
     &+&{\widehat \Gamma}^{(4)}_{AARA}{n_1\choose {-(1+n_1)}}
         {n_2\choose {-(1+n_2)}}{1\choose -1}{n_4\choose {-(1+n_4)}}
     \nonumber\\
     &+&{\widehat \Gamma}^{(4)*}_{AARA}{{(1+n_1)(1+n_2)(1+n_4)
         -n_1 n_2 n_4}\over {(1+n_3)-n_3}}
         {1\choose -1}{1\choose -1}{n_3\choose {-(1+n_3)}}{1\choose -1}
     \nonumber\\
     &+&{\widehat \Gamma}^{(4)}_{AAAR}{n_1\choose {-(1+n_1)}}
         {n_2\choose {-(1+n_2)}}{n_3\choose {-(1+n_3)}}{1\choose -1}
     \nonumber\\
     &+&{\widehat \Gamma}^{(4)*}_{AAAR}{{(1+n_1)(1+n_2)(1+n_3)
         -n_1 n_2 n_3}\over {(1+n_4)-n_4}}
         {1\choose -1}{1\choose -1}{1\choose -1}{n_4\choose {-(1+n_4)}}
     \nonumber\\
     &+&{\widehat \Gamma}^{(4)}_{RAAR}{1\choose -1}
         {n_2\choose {-(1+n_2)}}{n_3\choose {-(1+n_3)}}{1\choose -1}
     \nonumber\\
     &+&{\widehat \Gamma}^{(4)*}_{RAAR}{{(1+n_2)(1+n_3) - n_2 n_3}\over 
        {(1+n_1)(1+n_4) - n_1 n_4}}{n_1\choose {-(1+n_1)}}
        {1\choose -1}{1\choose -1}{n_4\choose {-(1+n_4)}}
     \nonumber\\
     &+&{\widehat \Gamma}^{(4)}_{RARA}{1\choose -1}
         {n_2\choose {-(1+n_2)}}{1\choose -1}{n_4\choose {-(1+n_4)}}
     \nonumber\\
     &+&{\widehat \Gamma}^{(4)*}_{RARA}{{(1+n_2)(1+n_4) - n_2 n_4}\over 
        {(1+n_1)(1+n_3) - n_1 n_3}}{n_1\choose {-(1+n_1)}}
        {1\choose -1}{n_3\choose {-(1+n_3)}}{1\choose -1}
     \nonumber\\
     &+&{\widehat \Gamma}^{(4)}_{RRAA}{1\choose -1}
         {1\choose -1}{n_3\choose {-(1+n_3)}}{n_4\choose {-(1+n_4)}}
     \nonumber\\
     &+&{\widehat \Gamma}^{(4)*}_{RRAA}{{(1+n_3)(1+n_4) - n_3 n_4}\over 
        {(1+n_1)(1+n_2) - n_1 n_2}}{n_1\choose {-(1+n_1)}}
        {n_2\choose {-(1+n_2)}}{1\choose -1}{1\choose -1}\, .
 \end{eqnarray}
Please note the much more symmetric structure of this equation
compared to Eq.~(\ref{G4vector}). A generalization of
Eqs.~(\ref{Ga3vect}) and (\ref{Ga4vect}) to $n\geq 5$ was recently
given in \cite{HWH98}. We emphasize once more that these
representations (in particular the vanishing of ${\widehat 
  \Gamma}^{(n)}_{R\dots R}$) rely heavily on the validity of the
generalized FDTs (or, equivalently, of the KMS condition). Similarly
simple expressions can thus not be expected to hold outside of thermal
equilibrium.

From Eqs.~(\ref{Gamarela}) and (\ref{hatgama4}) we can verify the
important symmetry relation
 \begin{equation}
 \label{GamaGama}
  {\widehat \Gamma}^{(n)}_{\Lambda_1\dots\Lambda_n}(k_1\dots k_n) =
     (-1)^{n+1}
     {{\prod \limits_{\{i|\Lambda_i=R\}} \Bigl( 1+ n(k^0_i)\Bigr)}\over 
      {\prod \limits_{\{i|\Lambda_i=A\}} n(k^0_i)}}
      {\widehat \Gamma}^{(n)*}_{{\bar \Lambda}_1\dots{\bar \Lambda}_n}
      (k_1\dots k_n)\, ,
 \end{equation} 
where ${\bar \Lambda}_i=A$ for $\Lambda_i=R$ and vice versa. A
similar relation was given in Ref.~\cite{Eijck}. 


\end{document}